# Molecular-Level Switching of Polymer/Nanocrystal Non-Covalent Interactions and Application in Hybrid Solar Cells

*Carlo Giansante,\* Rosanna Mastria, Giovanni Lerario, Luca Moretti, Ilka Kriegel, Francesco Scotognella, Guglielmo Lanzani, Sonia Carallo, Marco Esposito, Mariano Biasiucci, Aurora Rizzo, Giuseppe Gigli*

Dr. C. Giansante, G. Lerario, Dr. A. Rizzo, Prof. G. Gigli
Center for Biomolecular Nanotechnologies @UNILE, Istituto Italiano di Tecnologia, via Barsanti 1, 73010 Arnesano (LE), Italy
E-mail: carlo.giansante@iit.it

Dr. C. Giansante, R. Mastria, S. Carallo, M. Esposito, Dr. A. Rizzo, Prof. G. Gigli
NNL-CNR Istituto di Nanoscienze, via per Arnesano, 73100 Lecce, Italy
C.G. and R.M. contributed equally.

L. Moretti, Dr. I. Kriegel, Dr. F. Scotognella, Prof. G. Lanzani
Dipartimento di Fisica, Politecnico di Milano, P.zza L. da Vinci 32, 20133 Milano, Italy

Dr. M. Biasiucci
Dipartimento di Fisica G. Marconi, Università La Sapienza, 00185 Roma, Italy

Prof. G. Gigli
Dipartimento di Matematica e Fisica 'E. De Giorgi', Università del Salento, via per Arnesano, 73100 Lecce, Italy

**Abstract.**

Hybrid composites obtained upon blending conjugated polymers and colloidal semiconductor nanocrystals are regarded as attractive photoactive materials for optoelectronic applications. Here we demonstrate that tailoring nanocrystal surface chemistry permits to control non-covalent and electronic interactions between organic and inorganic components. We show that the pending moieties of organic ligands at the nanocrystal surface do not merely confer colloidal stability while hindering charge separation and transport, but drastically impact morphology of hybrid composites during formation from blend solutions. The relevance of our approach to photovoltaic applications is demonstrated for composites based on poly(3-hexylthiophene) and lead sulfide nanocrystals, considered as inadequate until this report, which enable the fabrication of hybrid solar cells displaying a power conversion efficiency that reaches 3 %. By investigating (quasi)steady-state and time-resolved photo-induced processes in the nanocomposites and their constituents, we ascertain that electron transfer occurs at the hybrid interface yielding long-lived separated charge carriers, whereas interfacial hole transfer appears hindered. Here we provide a reliable alternative aiming to gain control over macroscopic optoelectronic properties of polymer/nanocrystal composites by mediating their non-covalent interactions via ligands' pending moieties, thus opening new possibilities towards efficient solution-processed hybrid solar cells.

## 1. Introduction

Hybrid composites prepared by blends of conjugated polymers (CPs) and colloidal inorganic semiconductor nanocrystals (NCs) offer opportunities for the fabrication of large-area flexible optoelectronic devices exploiting cost-effective processing from solution-phase.[1-3] Indeed, research on hybrid nanocomposites promises to merge the potential advantages of both organic and inorganic components: namely, mechanical flexibility, low specific weight, and large absorption cross section of CPs[4] can be effectively combined with the small effective masses of charge carriers and good photochemical stability of NCs.[5] Among others, lead-chalcogenide NCs present the additional advantage of extending the light harvesting to the near-infrared (NIR) portion of the solar spectrum,[6] whereas the possibility of multiple exciton generation[7] and hot-carrier extraction[8] offers the opportunity to further improve the photon-to-charge-carrier conversion efficiency by reducing losses of high-energy carriers. In most cases, colloidal stability and consequent solution processability of NCs are guaranteed by ligands at the NC surface, which are already introduced during the synthetic procedure;[9] however, these pristine ligands are generally electrically insulating long-chain aliphatic compounds. The post-synthesis replacement of such pristine bulky ligands with short chemical species is therefore essential to promote charge transfer at the CP/NC interface and to confer good charge transport properties to the NC portion of the composite.[10,11]

Herein, we show that the pending moieties of NC ligands do not merely confer colloidal stability while hindering charge separation and transport, but exert a crucial role in determining non-covalent interactions between CPs and NCs thus affecting morphology and boundaries of hybrid CP/NC composites during formation from blend solutions. The consequent deep impact on optoelectronic properties is demonstrated for nanocomposites constituted by poly(3-hexylthiophene), commonly referred to as P3HT, and PbS NCs allowing their integration in solution-processed hybrid solar cells as photoactive materials, which could have been considered inappropriate until this report because corresponding devices have shown poor efficiency.[12,13]

To this aim, chemical modification of NC surface with short molecules which preserve colloidal stability is mandatory. However, the propensity of lead-chalcogenide NCs to oxidation, coalescence, and aggregation upon solution-phase ligand exchange has been a strong limitation to their integration in hybrid nanocomposites for optoelectronic applications. Indeed, with the exception of the incomplete pristine ligand exchange based on alkylamines,[14,15] no solution-phase ligand exchange method has been developed and exploited for blending lead-chalcogenide NCs and CPs. In this work we exploit a solution-phase ligand exchange approach that employs arenethiol molecules in presence of a base to produce strongly nucleophilic arenethiolate ligands which completely replace pristine oleate ligands on the PbS NC surface, while preserving good long-term solubility in chlorinated solvents[16,17] that are commonly used for dissolving CPs, including P3HT. Here we take advantage of such an intermediate step to subtly control non-covalent interactions between CP and NC counterparts (relevant chemical species employed in this work are shown in Figure 1a), permitting to tailor the morphology of the resulting nanocomposites for application in hybrid solar cells. Our approach thus represents a reliable alternative to grafting CPs to NCs via functional groups[18,19] or synthesizing NCs in CP matrix,[20,21] towards the nanometer size scale control of morphology and interfaces in solution-processed hybrid composites for optoelectronic applications.

## 2. Results and Discussion

### 2.1. Ligand-mediated morphology of the hybrid nanocomposites.

The effect exerted by organic ligands at the PbS NC surface clearly appears in electron microscopy (EM) images of the as-deposited nanocomposites, which reveal that as-synthesized oleate-capped PbS NCs[22] (henceforth referred to as PbS-Ol NCs) are evenly dispersed in the P3HT matrix, while arenethiolate-capped PbS NCs (hereafter referred to as PbS-ArS NCs) form well-defined nanometer scale domains preserving their peculiar cubic close-packing (compare panels b and c in **Figure 1**) already observed in mere PbS-ArS NC solids.[16] This striking difference can be attributed to the pending moieties of the NC ligands: hexyl side chains of P3HT may intercalate aliphatic chains of PbS-Ol NCs guaranteeing NC dispersion in the polymeric matrix, while strong ligand-ligand interactions between PbS-ArS NCs are likely to induce their close-packing thus promoting nanometer length scale separation of organic and inorganic phases. Noticeably, the peculiar ligand-ligand and ligand-polymer interactions which control the morphology of the nanocomposites are barely influenced by the relative ratio between P3HT and PbS NC components (detailed morphological characterization appears in Supporting Figures S2-19). The general applicability of our approach has been demonstrated employing other polymers bearing aliphatic and aromatic side chains blended with oleate- and arenethiolate-capped PbS or CdS[23] NCs (Supporting Figure S20).

The ligand-mediated arrangement of CPs and NCs in nanoscale domains is a key achievement towards the implementation of such hybrid nanocomposites in photovoltaic devices. As for conventional organic bulk-heterojunction solar cells,[24] the formation of interpenetrating domains provides both extended interfacial area for efficient exciton dissociation and percolating paths for effective charge carrier transport and extraction at the electrodes.[25] In order to exploit our molecular-level approach to control hybrid nanocomposite morphology in solar cell fabrication, we chemically and then thermally treated as-deposited nanocomposites to further improve charge transport properties of PbS NC and P3HT domains, respectively: 3-mercaptoproprionic acid (hereafter referred to as MPA) in acetonitrile solution was dispensed on the nanocomposites to replace Ol and ArS ligands[16,17] thus enhancing inter-NC coupling,[26] followed by mild annealing at 110 °C to promote P3HT inter-chain packing which increases carrier mobility.[27] EM imaging shows the effect exerted by post-deposition treatments on the morphology of the hybrid nanocomposites (see **Figure 2**). Such solid-phase treatments on P3HT/PbS-Ol NC composites promote the NC aggregation in submicrometric domains, suggesting partial demixing of the components; moreover, the large volume loss upon replacement of bulky Ol ligands with small MPA molecules results in extensive cracking of the nanocomposite (Figure 2a and 2c). Conversely, post-deposition treatments on P3HT/PbS-ArS NC composites do not significantly alter the nanometer scale NC domains already obtained during nanocomposite formation from blend solutions, whereas ensuring replacement of ArS ligands[16,17] and P3HT intrachain packing. In addition, such P3HT/PbS-ArS NC composites show a smooth profile and abated cracking (Figure 2b and 2d), which is of utmost importance to avoid short circuit and leakage currents in the photovoltaic devices.

## 2.2. Hybrid solar cells.

Hybrid nanocomposites for application in photovoltaic devices also demand for CP and NC components displaying staggered energy levels (also denoted as type-II band alignment) to promote separation of the photo-generated electron-hole pair in long-lived charge carriers. The synthetic control of NC diameter permits tuning of band-edge energies thanks to the quantum confinement effect: therefore, PbS NCs with diameter of about 3 nm have been synthesized[22] and used in order to maximize the absorption of solar light and the driving force for photo-induced charge carrier transfer with P3HT. Panel a of **Figure 3** illustrates a tentative energy level diagram of the frontier orbitals of P3HT[28] and PbS NCs,[29] suggesting the formation of a type-II bulk-heterojunction. Optimized devices were prepared by spin-casting from blend solutions of P3HT and PbS NCs at a 1:9 weight ratio on top of a transparent anode, followed by chemical treatment with MPA and thermal annealing (see Methods for fabrication details and Supporting Figures S21-22). A sketch of the photovoltaic device architecture is shown in Figure 3b.

As expected, the nanocomposite morphology largely impacts the current-voltage characteristics of the devices comprising P3HT/PbS-Ol NC and P3HT/PbS-ArS NC composites (shown in Figure 3c and 3d as grey and black lines, respectively) in dark and under solar-simulated illumination. Devices constituted by P3HT/PbS-ArS NC composites display power conversion efficiency (PCE) up to 3.0 % (maximum and average –in parentheses– recorded values are reported in **Table 1**), yielding an enhancement of three orders of magnitude compared to devices comprising P3HT/PbS-Ol NC composites (see Table 1). Such outstanding PCE values are two orders of magnitude larger than previous reports on the P3HT/PbS NC system[14,15] and about one order of magnitude larger than hybrid devices employing P3HT blended with other lead-chalcogenides NCs.[30,31]

The devices based on P3HT/PbS-ArS NC composites exhibit high reproducibility, with an average PCE value of 2.2 %, and are stable for at least one month after fabrication under inert atmosphere (see Supporting Figure S23). Dark $i$-V curves of these devices show good dark rectification behavior, low series (35 $\Omega cm^2$), and high shunt (29 $K\Omega cm^2$) area-normalized resistances, accounting for good charge transport properties and reduced internal current leakages. Under solar-simulated illumination, much larger short-circuit current density ($J_{sc}$) and open-circuit voltage ($V_{oc}$) values are obtained for devices based on P3HT/PbS-ArS NC composites compared to those values relative to P3HT/PbS-Ol NC composite based devices (see Table 1), together with an appreciable improvement of the fill factor (FF).

### 2.3. Photo-induced processes at the hybrid interface.

In order to ascertain whether the high efficiency in P3HT/PbS-ArS NC solar cells arises from a donor-acceptor heterojunction or is due to Schottky contacts at the anode,[32] we investigated the photo-induced processes in the hybrid nanocomposites by employing (quasi) steady-state and time-resolved spectroscopic techniques. The absorption spectrum of P3HT/PbS-ArS NC composite coincides with the sum of its components (panel a of **Figure 4** and Supporting Figure S30), indicating the absence of optically active ground state charge transfer. Upon excitation of both components in the blue spectral region, steady-state photoluminescence (PL) of P3HT is largely quenched upon blending with PbS-ArS NCs (> 95%, orange curve for neat P3HT compared to black spectrum of the hybrid composite in Figure 4b), clearly indicating the formation of additional deactivation pathways of the excited state localized on P3HT. The observed excited state dynamics (Figure 4c) suggest that the quenching of P3HT PL is limited by exciton diffusion which may occur in the organic domains before dissociation at the hybrid interface. Furthermore, almost complete quenching of PbS NC PL is observed upon post-deposition treatments of PbS-ArS NC solids (> 99%, dashed red curve for the PbS-ArS NCs as-casted from solution, compared to solid red curve representing the chemically and thermally treated PbS-ArS NC solid in Figure 4b), suggesting efficient exciton dissociation in the inorganic domains[33] already before blending.

On this basis, the differential transmittance at the first excitonic peak of PbS NCs (around 920 nm) observed in sub-picosecond transient absorption (TA) measurements of the post-deposition treated samples (both P3HT/PbS-ArS NC composite shown in panel a of **Figure 5** and PbS-ArS NC solid in Supporting Figure S32) can be related to dissociated charge carriers.[34] Noticeably, in presence of P3HT, the bleaching of PbS NC first exciton transition increases over 20 ps (compare black and red lines in Figure 5b), whereas bleaching of the ground state of P3HT ordered aggregates (at 610 nm, circles in Figure 5b) decreases within a comparable time scale (and faster than in the neat P3HT film, see Supporting Figure S33). These observations account a photo-induced electron transfer from P3HT to PbS NCs which is kinetically controlled by exciton diffusion at the hybrid interface, however indicating partial recombination on comparable time scale. At longer delays (> 100 ps), the dynamics at the first excitonic peak of PbS-ArS NCs in the composite and neat solid are similar (black and red lines in Figure 5b, respectively), pointing to an analogous recombination mechanism in both samples after the observed initial P3HT-to-NC electron transfer; moreover, P3HT lower excited state bleaching shows an increasing signal (at 610 nm, circles in Figure 5b) which follows the dynamics of the photo-induced absorption feature at 650 nm (triangles in Figure 5b) attributable prevalently to PbS-ArS NCs (see Supporting Figures S31-33).[35,36] In order to determine the presence of long-lived excited species, we employed quasi steady-state photo-induced absorption (PIA) spectroscopy: besides P3HT and PbS NC bleaching (at around 610 and 920 nm, respectively), PIA spectrum of the hybrid composite displays a strong broad feature reminiscent of P3HT delocalized polarons,[37] which gives a positive contribution to the differential absorbance signal centered around 700 nm displaying TA dynamics on the millisecond time scale (black spectrum and decay in Figure 5c-d, respectively). Such long-lived photo-excited species cannot be observed in the isolated components: indeed, PbS-ArS NC solids show different spectral features decaying in the microsecond time scale (red spectrum and decay in Figure 5c-d, respectively), whereas neat P3HT shows only weak polaron signal (orange decay in Figure 5d).

Conversely, upon selective excitation of PbS NCs, PIA spectrum of the hybrid composite shows little, if any, P3HT delocalized polaron signal superimposed to PbS NC photo-induced absorption feature (compare black and red spectra in panel a of **Figure 6**, respectively) implying that hole transfer from PbS NCs to P3HT is hindered. Despite previous reports accounting for NC-to-CP hole transfer,[38,39] a slight contribution of such photo-induced process to the generation of mobile charge carriers in hybrid composites has been argued.[40] In our nanocomposites however, the decay of PbS NC first excitonic peak bleaching is longer in presence of P3HT (Figure 6b), whereas the ground state of P3HT appears as depopulated upon PbS NC selective excitation (Figure 6a and 6c). These findings together with the weak P3HT polaron signals may imply that charges photo-generated in the PbS NC domains selectively accumulate at the hybrid interface, therefore inducing an electric field which shifts the energy of P3HT vibronic transitions rather than their apparent bleaching.

Further information on recombination of photo-generated charge carriers in the hybrid solar cells can be gathered by the incident light intensity ($P_{in}$) dependence of the $V_{oc}$.[41,42] The experimental data show that the slope of $V_{oc}$ vs. $\ln P_{in}$ equals the thermal voltage $k_B T/q$, where $k_B$ is the Boltzmann constant, T is the absolute temperature, and q is the elementary charge, which indicates that under open-circuit conditions most of the photo-generated charge carriers recombine at the hybrid interface according to second-order mechanism (see Supporting Figures S24-26). In order to correlate morphological and electronic properties of the P3HT/PbS-ArS NC composite, we performed Scanning Kelvin Probe Microscopy (SKPM) measurements which enable to obtain the surface potential of active layers.[43,44] SKPM maps reveal a larger excess of electrons on the surface of hybrid composite under white-light illumination when compared to its P3HT and PbS-ArS NC constituents and a much larger surface potential recover after illumination than in pure PbS-ArS NC solid (see Supporting Figures S27-29). All these findings account for a relatively low trap density giving a perspective to our molecular-level approach to gain control over the hybrid interface between CP and NC domains.

## 4. Conclusion

Here we have shown that the pending moieties of semiconductor NC ligands dictate the morphology of hybrid nanocomposites with CPs exploiting peculiar ligand-ligand and ligand-polymer interactions. Arenethiolate ligands promote phase segregation at the nanometer size scale between PbS NCs and P3HT components, which is not significantly altered by post-deposition chemical[25] and thermal[26] treatments, yielding a hybrid nanocomposite that displays distinct photoconductive inorganic and organic domains with dimension of about 10÷100 nm, which is comparable to the charge carrier and exciton diffusion lengths in solids of pure PbS NCs[45] and P3HT,[46] respectively. This nanoscale arrangement allows electron-hole pairs photo-generated in the P3HT matrix to reach the extended hybrid interfacial area where they can separate in long-lived charge carriers which are expected to get to the electrodes via eventual percolating paths, albeit partial recombination occurs. The overall result is an impressive thousand-fold enhancement of the PCE in devices incorporating P3HT/PbS-ArS NC composites compared to the efficiency obtained employing composites based on P3HT and as-synthesized PbS-Ol NCs.

More broadly, we provide a remarkable example of the opportunities offered by NC surface chemistry for the nanoscale tuning of the macroscopic optoelectronic properties of hybrid composites. The obtainment of colloidally stable NCs is crucial to this purpose and, despite the evident benefits of solid-phase ligand exchange on the photovoltaic performances of hybrid composites based on narrow bandgap materials,[47-49] we emphasize the possibility of dictating the nanocomposite features during its formation from blend solutions via chemical modification of NC surface in solution-phase. Indeed, the pending moieties of the NC ligands do not merely hinder conductivity, but represent a simple yet powerful tool to gain control over the macroscopic properties of hybrid composites by mediating the morphology and boundaries at the nanometer size scale. Such an approach may represent a reliable alternative to grafting CPs to NCs via functional groups[19,20] or synthesizing NCs in the CP matrix,[21,22] thus contributing to suggest novel pathways towards efficient CP/NC solar cells, which still lags behind the efficiencies obtained for devices based on pure polymer[50] and nanocrystal[51] active layers.

## 5. Experimental Section.

*Materials.* All chemicals were of the highest purity available unless otherwise noted and were used as received. Lead oxide (99.999%), oleic acid (technical grade 90%), 1-octadecene (technical grade 90%), bis(trimethylsylil)sulfide (synthesis grade), 4-methylbenzenethiol (ArSH, 98%), 3-mercaptoproprionic acid (MPA, $\geq$ 99%), and regioregular poly(3-hexylthiophene-2,5-diyl) (P3HT, average molecular weight 54,000÷75,000 u) were purchased from Sigma-Aldrich. Tri-n-octylphosphine (TOP, 97) was purchased from Strem Chemicals. Triethylamine (TEA, $\geq$ 99.5%) was purchased from Fluka. All solvents were anhydrous and were used as received. Acetone (99.8%) was purchased from Merck. Acetonitrile (99.8%), chloroform (99.8%), o-dichlorobenzene (99%), hexane (95%), methanol (99.8%), and toluene (99.8%) were purchased from Sigma-Aldrich.

*NC synthesis and solution-based ligand exchange procedure.* Colloidal PbS-Ol NCs with diameter of about 3 nm were synthesized in a three-neck flask connected to a standard Schlenk line setup under oxygen- and water-free conditions using lead(II)-oleate and bis(trimethylsylil)sulfide at a 2:1 molar ratio through a slightly modified well-established procedure.[22] In a typical synthesis, 2 mmol of PbO (450 mg) was mixed with 4 mmol (1140 mg) of oleic acid and 10 g of 1-octadecene. The mixture was vigorously stirred and deaerated through repeated cycles of vacuum application and purging with nitrogen. Then, the mixture was heated to above 100 °C to allow dissolution of PbO until the solution became colorless and optically clear, indicating the formation of lead(II)-oleate complex. The solution was cooled at 80 °C and repeatedly subjected to vacuum in order to remove water formed upon lead(II)-oleate complex formation. The solution was then heated again under nitrogen flow and stabilized at 110°C. At this point 1 mmol of bis(trimethylsylil)sulfide (210 μL) in 2 mL of TOP was swiftly injected. The heating mantle was immediately removed and the resulting black colloidal solution was allowed to naturally cool to room temperature. After the synthesis, PbS-Ol NCs were transferred to a nitrogen-protected glove box. The NCs were precipitated using excess acetone, centrifuged at 4000 rpm for 7 min and then redissolved in toluene. Centrifugation at 4000 prm for 5 minutes and filtration through a 0.2 μm polytetrafluoroethylene (PTFE) membrane were carried out to discard insoluble products and possible agglomerates. Additional precipitation-redissolution cycles were applied to remove any excess unbound surfactants. A 1 mM NC solution was prepared and stored at room temperature in a glove box for producing PbS-ArS NCs.[16,17] Solution-phase ligand exchange on PbS-Ol NCs (reaction scheme is shown in Supporting Figure S1) was achieved by adding a slight excess of p-methylbenzenethiol/triethylamine, compared to that calculated by spectrophotometric titration, to a 1 mM solution at room temperature in oxygen free atmosphere. In a typical solution ligand exchange procedure, 300 equivalents of p-methylbenzenethiol/triethylamine were added to a 1 mM PbS-Ol NCs solution. The mixture was precipitated with hexane and methanol, centrifuged and redispersed in dichlorobenzene and/or chloroform. The purification procedure was repeated twice. Final dispersion is thus centrifuged and passed through a 0.2 μm PTFE membrane to discard insoluble products and eventual agglomerates. The yield of the solution ligand exchange process, estimated from ICP-AES measurements, is up to 90 %. Complete ligand exchange was demonstrated by the disappearance of carboxylate peaks in FTIR spectra and by the presence of features characteristic of ArS; furthermore the optical absorption spectra of PbS-ArS NCs were superimposable, regardless of purification procedure, so confirming that ArS efficiently displaced Ol ligands, did not induce aggregation of PbS NCs and preserved good long-term colloidal stability. 1 mM PbS-ArS NC solutions were stable up to several weeks.

*Morphological characterization.* Low-resolution transmission EM images were recorded with a Jeol Jem 1011 microscope operated at an accelerating voltage of 100 kV. Samples for analysis were prepared by spin-coating the blend solutions on glass/PEDOT:PSS substrates, then floated off the substrate onto the surface of a water bath, and transferred to carbon-coated Cu grids. Scanning EM images of the samples were recorded using Carl Zeiss Auriga40 Crossbeam instrument, in high-vacuum and high-resolution acquisition mode, equipped with Gemini column and an integrated high efficiency In-lens SE (secondary electrons) detector. The applied acceleration voltage was 20 kV, with an additional 8kV booster voltage to increase signal to noise ratio. Samples for analysis were prepared by spin-coating the blend solutions on Si substrates.

*Device fabrication and characterization.* Patterned ITO-coated glass substrates (Visiontek) were cleaned using TL1 solution ($H_2O$/$NH_4OH$/$H_2O_2$ 5:1:1) and covered with a 40 nm thick layer of poly(3,4-ethylenedioxythiophene)/poly(styrenesulfonate) (PEDOT:PSS, Clevios P 4083). Concentrated PbS NC solution (90 mg/ml in dichlorobenzene/chloroform 3:2) was blended with a P3HT solution (13 mg/ml in dichlorobenzene) at a 9:1 weight ratio and spin-casted at 1500 rpm for 60 sec yielding ~ 100 nm thick nanocomposite films. These films were treated with 15 mM acetonitrile solution of MPA and spin-coated at 1500 rpm for 30 sec, then applying two washing steps with mere acetonitrile. An annealing step at 110 °C for 30 min was then performed. The thin films were kept under vacuum over night. LiF (0.6 nm) and Al (130 nm) layers were deposited on top of the nanocomposites by thermal evaporation. The device area (4 mm$^2$) was determined by the overlap of the ITO and Al electrodes and accurately measured using an optical microscope. The current–voltage characteristics were determined using an Air Mass 1.5 global (AM 1.5 G) solar simulator (Spectra Physics Oriel150W) with an irradiation intensity of 100 mW cm$^{-2}$ and recorded with a Keithley 2400 source meter. The spectral response of the short-circuit current was measured by using a 150 W Xe arc light source coupled to a monochromator, detected by a Si photodiode and recorded by a dual channel Merlin lock-in amplifier.

*Spectroscopic measurements.* The optical absorption spectra of the nanocomposites and their components were recorded with Varian Cary 5000 and JASCO UV-Vis-NIR spectrophotometers. Steady-state PL and PIA spectra and time-resolved PL and TA decays were recorded using homemade setups. In PIA and TA measurements, the probe light was generated by a 250 W tungsten-halogen lamp (Thermo Oriel) passing through a monochromator before being focused on the sample. The light transmitted through the sample was focused onto a second monochromator and detected by a Si photodiode connected to a current amplifier (Femto DLPCA200). In steady-state PL and PIA measurements, pump excitation is provided by continuous wave lasers: Spectra Physics Cyan 488 nm and Spectra Physics 3900S tunable Ti:Sapphire (tuned at 720 nm) were employed, with excitation power density of about 150 and 200 mWcm$^{-2}$, respectively, and modulation frequency of 200÷450 Hz. In TA measurements, an Innolas SplithLight Compact 200 OPO pulsed laser (~4 ns pulse, tuned at 488 nm) was used with an excitation intensity of approximately 200 mWcm$^{-2}$ at a pulse rate of about 10 Hz. The excitation light for time-resolved PL measurements was provided by a Mai Tai Spectra Physics pulsed laser (<100 fs pulse; 78 MHz repetition rate) tuned at 900 nm and passing through a frequency doubler; the excitation power is of about 200 mWcm$^{-2}$. The signals were recorded by a lock-in amplifier (SR 830 Stanford Research System) in steady-state PL and PIA measurements, by an oscilloscope (Tektronix TDS 2024C) in TA experiments, and by a streak camera (Hamamatsu C5680) coupled through a monochromator in time-resolved PL measurements. The laser system employed for ultrafast transient absorption was based on a Ti-Sapphire chirp pulse amplified source, with maximum output energy of about 800 μJ, 1 kHz repetition rate, central wavelength of 800 nm and pulse duration of about 180 fs. Excitation pulses at 400 nm were obtained by doubling the fundamental frequency in a β-Barium borate (BBO) crystal while other pump photons at different wavelength were generated by non-collinear optical parametric amplification in BBO, with pulse duration around 100 fs. Pump pulses were focused in a 175 μm diameter spot. Probing was achieved in the visible and near IR region by using white light generated in a thin sapphire plate. Chirp-free transient transmission spectra were collected by using a fast optical multichannel analyzer (OMA) with dechirping algorithm. The measured quantity is the normalized transmission change, ΔT/T. Excitation energy has been kept around 20 nJ, in order to get around 20 μJ/cm$^2$ excitation fluences. All measurements were performed at room temperature on sealed samples prepared under inert atmosphere.

*Scanning Kelvin Probe Microscopy measurements.* SKPM images were acquired with a commercial non-contact atomic force microscopy AFM system (Bruker-AXS) MultiMode AFM with Nanoscope V controller operating in Lift Mode (typical lift height 20 nm) by using a silicon tips with PtIr coating (SCM-PIT) with k ≈ 3 N/m (tip radius ≈ 20 nm, and resonant frequency ≈ 80 kHz). Surface potential maps were acquired in dark and under illumination by means of a halogen lamp, with an estimated fluence of 0.45 kW/m$^2$. The SKPM technique is performed by means of a non-contact AFM enables to measure the difference between the tip potential and the local surface potential with a lateral resolution below 100 nm and a potential distribution of 10 mV.

**References.**

[1]    E. Holder, N. Tessler,; A. L. Rogach, *J: Mater. Chem.* **2008,** *18* (10), 1064.

[2]    P. Reiss, E. Couderc, J. De Girolamo, A. Pron,. *Nanoscale* **2011,** *3* (2), 446.

[3]    F. Gao, S. Ren, J. Wang, *Energy Environ. Sci.* **2013,** *6* (7), 2020.

[4]    M. Kaltenbrunner, M. S. White, E. D. Głowacki, T. Sekitani, T. Someya, N. S. Sariciftci, S. Bauer, S., *Nat. Commun.* **2012,** *3*, 770.

[5]    D. V. Talapin, J.-S. Lee, M. V. Kovalenko, E. V. Shevchenko, *Chem. Rev.* **2009,** *110* (1), 389.

[6]    F. W. Wise, *Acc. Chem. Res.* **2000,** *33* (11), 773.

[7]    A. J. Nozik, M. C. Beard, J. M. Luther, M. Law, R. J. Ellingson, J. C. Johnson, *Chem. Rev.* **2010,** *110* (11), 6873.

[8]    W. A. Tisdale, K. J. Williams, B. A. Timp, D. J. Norris, E. S. Aydil, X.-Y. Zhu, *Science* **2010,** *328* (5985), 1543.

[9]    Y. Yin, A. P. Alivisatos, *Nature* **2005,** *437* (7059), 664-70.

[10]   N. C. Greenham, X. Peng, A. P. Alivisatos, *Phys. Rev. B* **1996,** *54* (24), 17628.

[11]   W. U. Huynh, J. J. Dittmer, W. C. Libby, G. L. Whiting, A. P. Alivisatos, *Adv. Funct. Mater.* **2003,** *13* (1), 73.

[12]   S. Günes, K. P. Fritz, H. Neugebauer, N. S. Sariciftci, S. Kumar, G. D. Scholes, *Sol. Energy Mater. Sol. Cells* **2007,** *91* (5), 420.

[13]   J. Seo, S. J. Kim, W. J. Kim, R. Singh, M. Samoc, A. N. Cartwright, P. N. Prasad, *Nanotechnology* **2009,** *20* (9), 095202.

[14]   S. A.McDonald, G. Konstantatos, S. Zhang, P. W. Cyr, E. J. D. Klem, L. Levina, E. H. Sargent, *Nature Mater.* **2005,** *4* (2), 138.

[15]  K. M. Noone, E. Strein, N. C. Anderson, P.-T. Wu, S. A. Jenekhe, D. S.Ginger, *Nano Lett.* **2010,** *10* (7), 2635.

[16]  C. Giansante, L. Carbone, C. Giannini, D. Altamura, Z. Ameer, G. Maruccio, A. Loiudice, M. R. Belviso, P. D. Cozzoli, A. Rizzo, A. Gigli, G., *J. Phys. Chem. C* **2013,** *117* (25), 13305.

[17]  C. Giansante, L. Carbone, C. Giannini, D. Altamura, Z. Ameer, G. Maruccio, A. Loiudice, M. R. Belviso, P. D. Cozzoli, A. Rizzo, A. Gigli, G., *Thin Solid Films* **2014**, *560*, 2.

[18]  J. Liu, T. Tanaka, K: Sivula, A. P. Alivisatos, J. M. J. Fréchet, *J. Am. Chem. Soc.* **2004,** *126* (21), 6550.

[19]  H. Skaff, K. Sill, T. Emrick, *J. Am. Chem. Soc.* **2004,** *126* (36), 11322.

[20]  H. C. Leventis, S. P. King, A. Sudlow, M. S. Hill, K. C. Molloy, S. A. Haque, *Nano Lett.*, **2010**, 10 (4), 1253.

[21]  A. A. R.Watt, D. Blake, J. H. Warner, E. A. Thomsen, E. L. Tavenner, H. Rubinsztein-Dunlop, P. Meredith, *J. Phys. D: Appl. Phys.* **2005,** *38* (12), 2006.

[22]  M. A. Hines, G. D. Scholes, *Adv. Mater.* **2003,** *15* (21), 1844.

[23]  W. W. Yu, X. Peng, *Angew. Chem. Int. Ed.* **2002,** *41* (13), 2368.

[24]  S. Günes, H. Neugebauer, N. S. Sariciftci, *Chem. Rev.* **2007,** *107* (4), 1324.

[25]  M. Graetzel, R. A. J. Janssen, D. B. Mitzi, E. H. Sargent, *Nature* **2012,** *488* (7411), 304-312.

[26]  A. G. Pattantyus-Abraham, I. J. Kramer, A. R. Barkhouse, X. Wang, G. Konstantatos, R. Debnath, L. Levina, I. Raabe, M. K. Nazeeruddin, M. Grätzel, E. H. Sargent, *ACS Nano* **2010,** *4* (6), 3374.

[27]  Y. Kim, S. Cook, S. M: Tuladhar, S. A. Choulis, J. Nelson, J. R.  Durrant, D. D. C. Bradley, M. Giles, I. McCulloch, C.-S. Ha, M. Ree, *Nature Mater.* **2006,** *5* (3), 197.

[28]  Z. Xu, L.-M. Chen, M.-H. Chen, G. Li, Y. Yang, Y, *Appl. Phys. Lett.* **2009,** *95* (1), 013301.

[29]  P. R. Brown, D. Kim, R. R. Lunt, N. Zhao, M. G. Bawendi,J. C. Grossman, V. Bulovic, *ACS Nano* **2014,** *8* (6), 5863.

[30]  D. Cui, J. Xu, T. Zhu, G. Paradee, S. Ashok, M. Gerhold, *Appl. Phys. Lett.* **2006,** *88* (18), 183111.

[31]  M. Nam, S. Kim, T: Kim, S.-W. Kim, K.-K. Lee, *Appl. Phys. Lett.* **2011,** *99* (23), 233115.

[32]  j. M. Luther, M. Law, M. C. Beard, Q. Song, M. O. Reese, R. J.  Ellingson, A. J. Nozik, *Nano Lett.*, **2008**, *8* (10), 3488.

[33]  J. J. Choi, J. Luria, B.-R. Hyun, A. C. Bartnik, L. Sun, Y.-F. Lim, J. A. Marohn, F. W. Wise, T. Hanrath, *Nano Lett.* **2010**, *10*, 1805.

[34]  J. Gao, C. S. S. Sandeep, J. M.Schins, A. J. Houtepen, L. D. A. Siebbeles, *Nat. Commun.*, **2013**, *4*, 2329.

[35]  F. Gesuele, M. Y. Sfeir, W. K. Koh, C. B. Murray, T. F. Heinz, C. W. Wong *Nano Lett.*, **2012**, *12* (6), 2658.

[36]  M. T. Trinh, M. Y. Sfeir, J. J. Choi, J. S. Owen, X.-Y. Zhu, *Nano Lett.*, **2013**, *13* (12), 6091.

[37]  J. Guo, H. Ohkita, H. Benten, S. Ito, *J. Am. Chem. Soc.*, **2010**, *132* (17), 6154.

[38]  E. Strein, D. W. deQuilettes, S. T. Hsieh, A. E. Colbert, D. S. Ginger, *J. Phys. Chem. Lett.*, **2014**, *5* (1), 208.

[39]  M. Panahandeh-Fard, J. Yin, M. Kurniawan, Z. Wang, G. Leung, T. C. Sum, C. Soci *J. Phys. Chem. Lett.*, **2014**, *5* (7), 1144.

[40]  F. S. F. Morgenstern, A. Rao, M. L. Böhm, R. J. K. Kist, Y. Vaynzof, N. C. Greenham, *ACS Nano*, **2014**, (2), 1647.

[41]  S. R. Cowan, A. Roy, A. J. Heeger, *Phys. Rev. B* **2010,** *82* (24), 245207.

[42] M. M. Mandoc, F. B. Kooistra, J. C. Hummelen, B. de Boer, P. W. M. Blom, *Appl. Phys. Lett.* **2007,** *91* (26), 263505.

[43] H. Hoppe, T. Glatzel, M. Niggmann, W. Schwinger, F. Shaeffer, A. Hinsch, M. C. Lux-Steiner, N. S. Sariciftci, *Thin Solid Film* **2006**, *511* (512), 587.

[44] G. Grancini, M. Biasiucci, R. Mastria, F. Scotognella F. Tassone, D. Polli, G. Gigli, G. Lanzani, *J. Phys. Chem. Lett.* **2012**, *3* (4), 517.

[45] D. Zhitomirsky, O. Voznyy, S. Hoogland, E. H. Sargent, *ACS Nano* **2013,** *7* (6), 5282.

[46] P. E. Shaw, A. Ruseckas, I. D. W. Samuel, *Adv. Mater.* **2008,** *20* (18), 3516.

[47] J. Seo, M. J. Cho, D. Lee, A. N. Cartwright, P. N. Prasad, *Adv. Mater.* **2011,** *23* (34), 398.

[48] C. Piliego, M. Manca, R. Kroon, M. Yarema, K. Szendrei, M. R. Andersson, W. Heiss, M. A. Loi, *J. Mater. Chem.* **2012,** *22* (46), 24411.

[49] Z. Liu, Y. Sun, J. Yuan, H. Wei, X. Huang, L: Han, W. Wang, H. Wang, W. Ma, *Adv. Mater.* **2013**, *25*, 5772.

[50] R. Zhou, R. Stalder, D. Xie, W. Cao, Y. Zheng, Y. Yang, M. Plaisant, P. H. Holloway, K. S. Schanze, J. R. Reynolds, J. Xue, *ACS Nano* **2013,** *7* (6), 4846.

[51] M Eck, C. Van Pham, S. Zufle, M. Neukom, M. Sessler, D. Scheunemann, E. Erdem, S. Weber, H. Borchert, B. Ruhstaller,M. Kruger, *Phys. Chem. Chem. Phys.* **2014,** *16*, 12251.

[52] L. Dou, J. You, J. Yang, C.-C. Chen, Y. He, S. Murase, T. Moriarty, K. Emery, G. Li, Y. Yang, *Nature Photon.* **2012,** *6* (3), 180.

[53] A. H. Ip, S. M. Thon, S. Hoogland, O. Voznyy, D. Zhitomirsky, R. Debnath, L. Levina, L. R. Rollny, G. H. Carey, A. Fischer, K. W. Kemp, I. J. Kramer, Z. Ning, A. J. Labelle, K. W. Chou, A. Amassian, E. H. Sargent, *Nature Nanotech.* **2012,** *7* (9), 577.

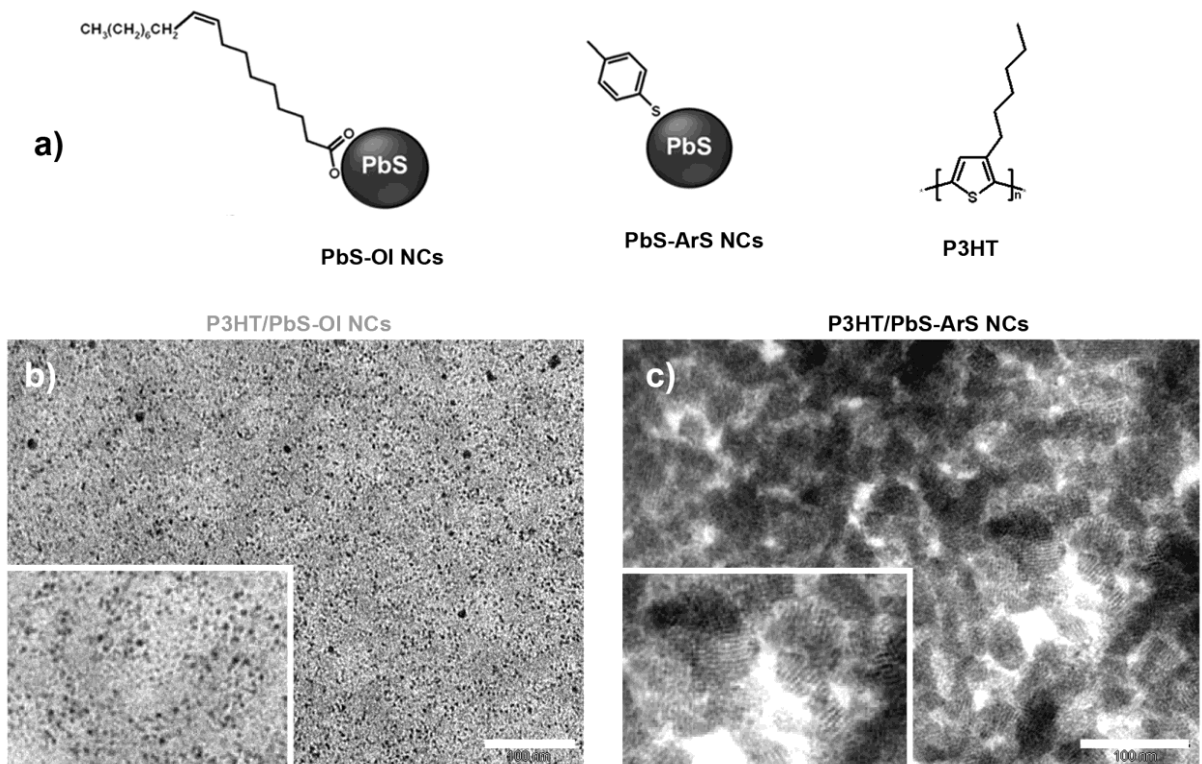

**Figure 1.** *Morphology of the as-deposited hybrid composites*. a) Schematic representation of the chemical structures and related acronyms of the relevant chemical species used in this work. b,c) Transmission EM images of the as-deposited nanocomposites. Insets in the bottom left corner of these panels provide a magnified view of the nanocomposites. Size scale bar appears in the bottom right corners and corresponds to 100 nm.

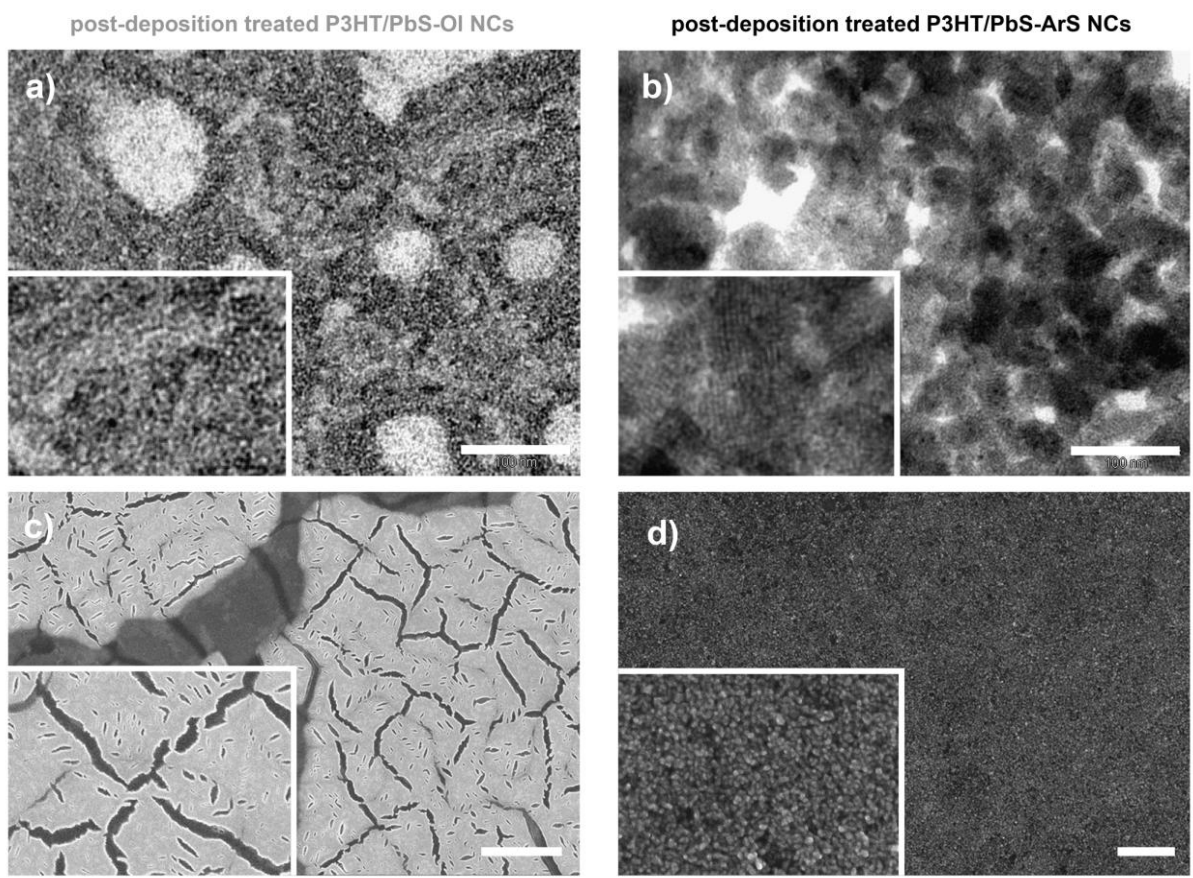

**Figure 2.** *Morphology of the post-deposition treated hybrid composites*. a,b) Transmission and (c,d) scanning EM images of the nanocomposites based on (a,c) P3HT/PbS-Ol NCs and (b,d) P3HT/PbS-ArS NCs after chemical and thermal post-deposition treatments. Insets in the bottom left corner of each panel provide a magnified view of the nanocomposites. Size scale bar appears in the bottom right corner of each panel and corresponds to (a,b) 100 nm or (c,d) 2 µm.

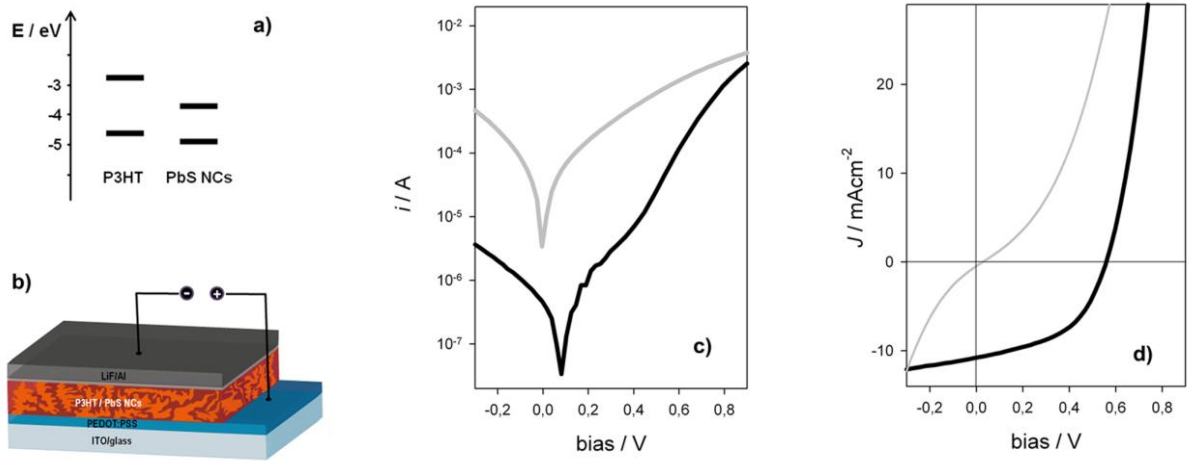

**Figure 3.** *Solar cells based on the hybrid composites*. a) Tentative energy level diagram for P3HT and PbS NCs according to references 28 and 29, respectively. b) Drawing of the hybrid bulk-heterojunction device architecture. c) Dark *i*-V characteristics of devices based on P3HT/PbS-Ol NCs (grey line) and P3HT/PbS-ArS NCs (black line). d) *J*-V characteristics under 100 mWcm$^{-2}$ solar-simulated AM1.5G light illumination of the bulk-heterojunction solar cells prepared from P3HT/PbS-Ol NC (grey line) and P3HT/PbS-ArS NC (black line) composites.

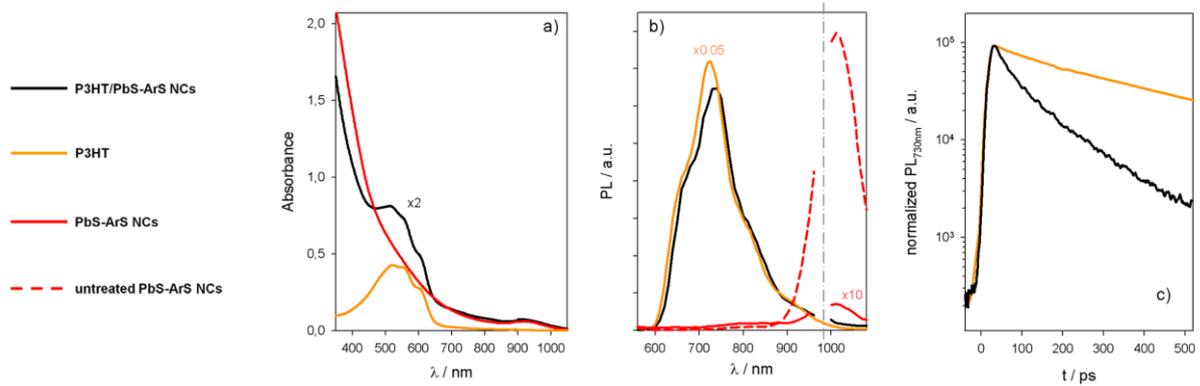

**Figure 4.** *Exciton dissociation in the hybrid composites*. Legend appears in the top left corner. a)Absorption spectra of the P3HT/PbS-ArS NC composite (black line) and of its components (neat P3HT, orange line, and PbS-ArS NC film, red line). b) Steady-state PL spectra ($\lambda_{ex}$ = 488 nm) of mere P3HT (orange line), P3HT/PbS-ArS NC composite (black line), and PbS-ArS NC solids (red lines) as-casted from solution (dashed line) and chemically and thermally treated; grey line indicates the second harmonic of the excitation light. All PL spectra have been normalized for their absorption at the excitation wavelength. c) Time-resolved PL ($\lambda_{ex}$ = 450 nm; $\lambda_{em}$ = 720÷740 nm) of mere P3HT (orange line) and P3HT/PbS-ArS NC composite (black line).

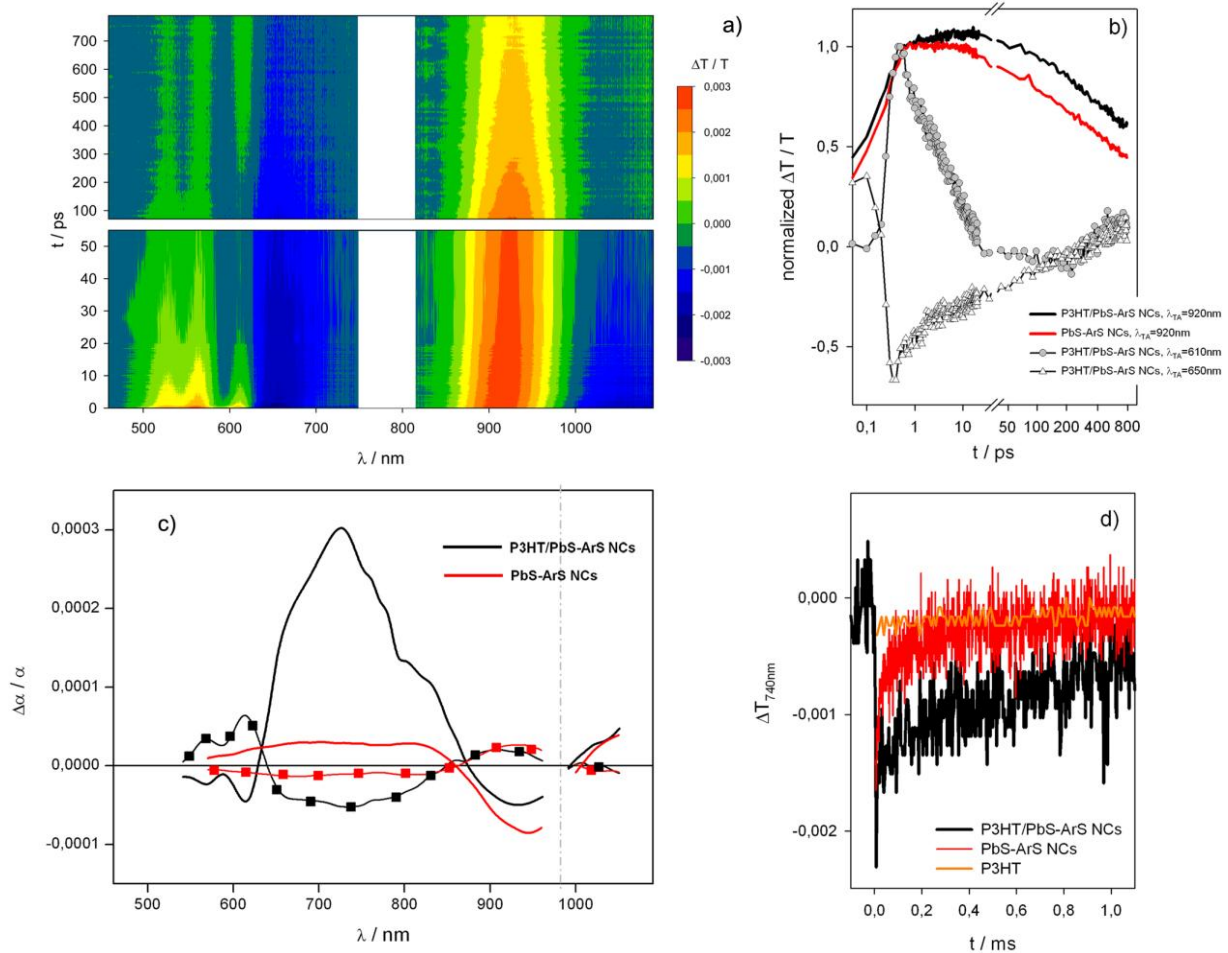

**Figure 5.** *Electron transfer in the hybrid composites upon P3HT co-excitation.* a) TA spectrum of P3HT/PbS-ArS NC composite ($\lambda_{pump}$ = 400 nm at a pulse irradiance of 20 $\mu$J/cm$^2$); the missing part of the spectrum is due to the fundamental of the Ti:Sapphire amplified system. b) Normalized TA kinetics ($\lambda_{pump}$ = 400 nm) of P3HT/PbS-ArS NC composite (black lines) probed at the bleaching of PbS NC first excitonic peak ($\lambda_{probe}$ = 920 nm, solid line), at the bleaching of P3HT 0-0 vibronic transition ($\lambda_{probe}$ = 610 nm, circle line), and at the photo-induced absorption of PbS NCs ($\lambda_{probe}$ = 650 nm, triangle line; a vertical offset has been applied for clarity); red line is the normalized TA decay of PbS-ArS NC solid ($\lambda_{pump}$ = 400 nm, $\lambda_{probe}$ = 920 nm). c) PIA spectra ($\lambda_{pump}$ = 488 nm at a fluence of about 75 mW/cm$^2$) of P3HT/PbS-ArS NC composite (black lines) and PbS-ArS NC solid (red lines). In-phase channel PIA spectra are represented by solid lines, while squared lines represent the out-of-phase channel of PIA spectra); grey line represents the second harmonic of $\lambda_{pump}$. d) TA kinetics ($\lambda_{pump}$ = 488 nm; $\lambda_{probe}$ = 740 nm) of P3HT/PbS-ArS NC composite (black line), neat P3HT (orange line) and PbS-ArS NC solid (red line).

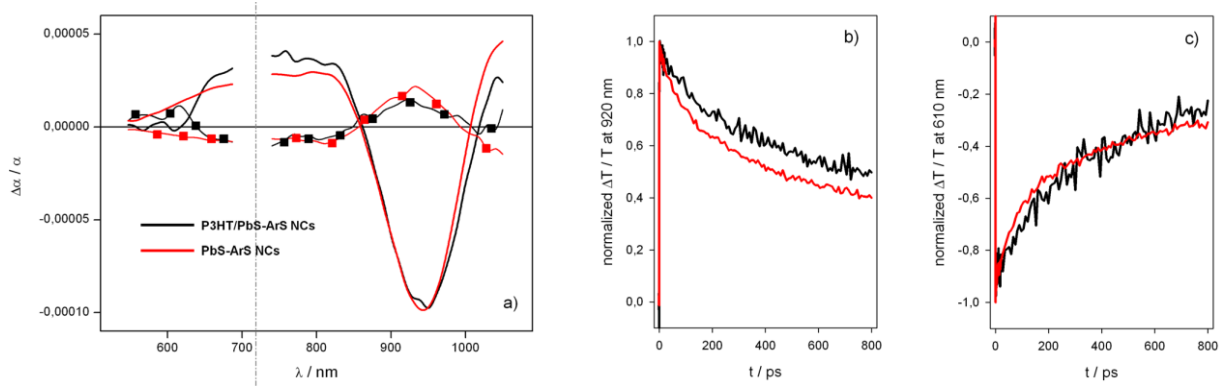

**Figure 6.** *Hindered hole transfer in the hybrid composites upon PbS-ArS NC selective excitation.* a) PIA spectra ($\lambda_{pump}$ = 720 nm at a fluence of about 200 mW/cm$^2$) of the P3HT/PbS-ArS NC composite (black lines) and of PbS-ArS NC film (red lines). In-phase channel PIA spectra are represented by solid lines, while squared lines represent show the out-of-phase channel of PIA spectra; grey line indicates $\lambda_{pump}$. b,c) Normalized TA kinetics ($\lambda_{pump}$ = 700 nm at a fluence of about 20 µJ/cm$^2$) of P3HT/PbS-ArS NC composite (black lines)and PbS-ArS NC solid (red lines) probed at 920 nm and 610 nm (panels b and c, respectively).

**Table 1.** *Summary of the photovoltaic parameters of devices based on the hybrid nanocomposites.* $J_{sc}$, $V_{oc}$, FF, and PCE stand for short-circuit current density, open circuit voltage, fill factor, and, power conversion efficiency, respectively. Maximum values (average values and one standard deviation are reported in parentheses) are obtained by characterizing 16 devices fabricated employing nanocomposites from the same batch of synthesized and ligand exchanged PbS NCs.

| Active layer | $J_{sc}$ / mAcm$^{-2}$ | $V_{oc}$ / V | FF / % | PCE / % |
|---|---|---|---|---|
| P3HT/PbS-Ol NCs | 0.43 | 0.026 | 25 | 0.0028 |
| **P3HT/PbS-ArS NCs** | **10.8** **(10.4 ± 0.88)** | **0.56** **(0.46 ± 0.08)** | **50** **(46 ± 2.7)** | **3.0** **(2.2 ± 0.35)** |

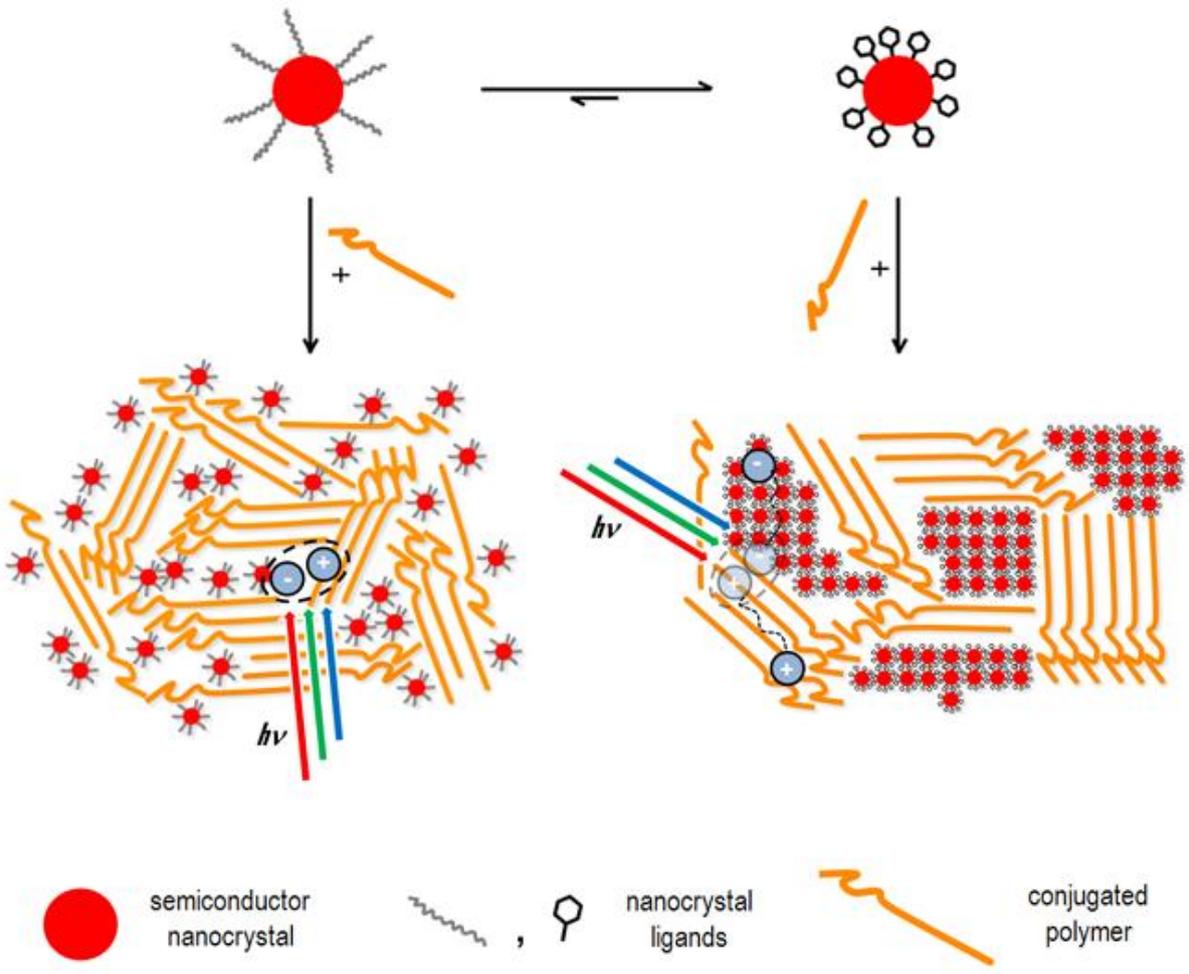

semiconductor nanocrystal , nanocrystal ligands , conjugated polymer

# Supporting Information

**Molecular-Level Switching of Polymer/Nanocrystal Non-Covalent Interactions and Application in Hybrid Solar Cells**

*Carlo Giansante,\* Rosanna Mastria, Giovanni Lerario, Luca Moretti, Ilka Kriegel, Francesco Scotognella, Guglielmo Lanzani, Sonia Carallo, Marco Esposito, Mariano Biasiucci, Aurora Rizzo, Giuseppe Gigli*

***PbS NC synthesis and solution-based ligand exchange procedure***.

Colloidal PbS-Ol NCs with diameter of about 3 nm were synthesized using lead(II)-oleate and bis(trimethylsylil)sulfide at a 2:1 molar ratio following a well established procedure.[1] Colloidal PbS-ArS NCs were obtained by adding a slight excess of p-methylbenzenethiol/triethylamine to PbS-Ol NCs, as already reported.[2] In a typical procedure, up to 300 equivalents of p-methylbenzenethiol/triethylamine were added to a 1 mM PbS-Ol NCs in dichlorobenzene solution. The mixture was precipitated with hexane and methanol, centrifuged and redispersed in dichlorobenzene or chloroform. The purification procedure has been repeated twice.

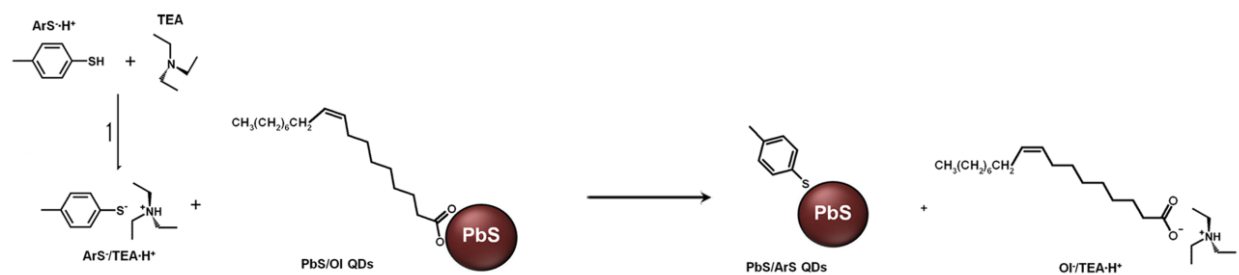

**Figure S1.** Solution-phase ligand exchange reaction scheme.

*Morphological characterization of the hybrid nanocomposites.*

*TEM images of P3HT/PbS-Ol NC composites*

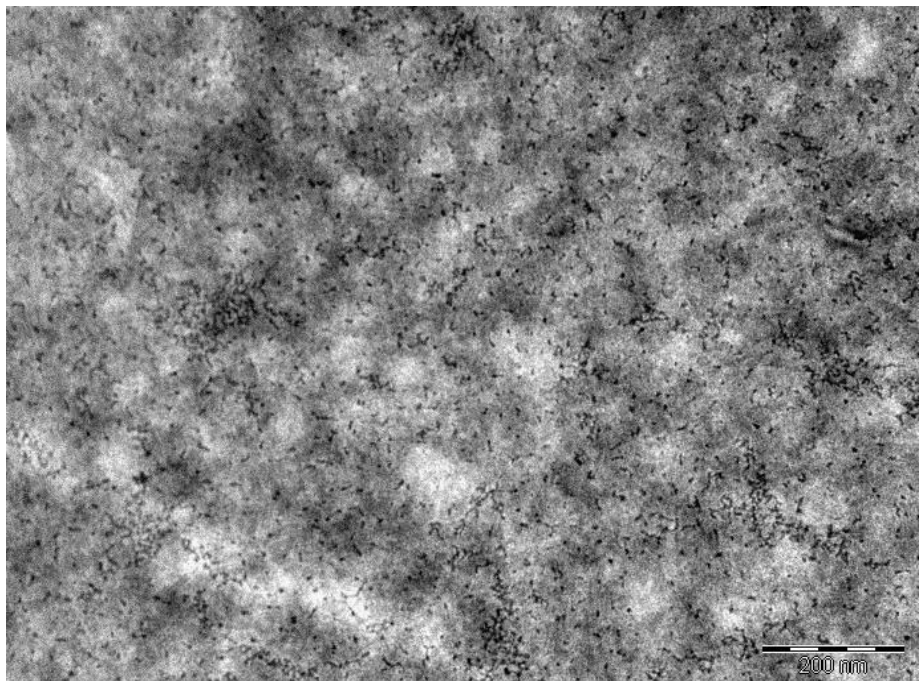

**Figure S2.** TEM image of the composite P3HT/PbS-Ol NCs 50:50 w/w.

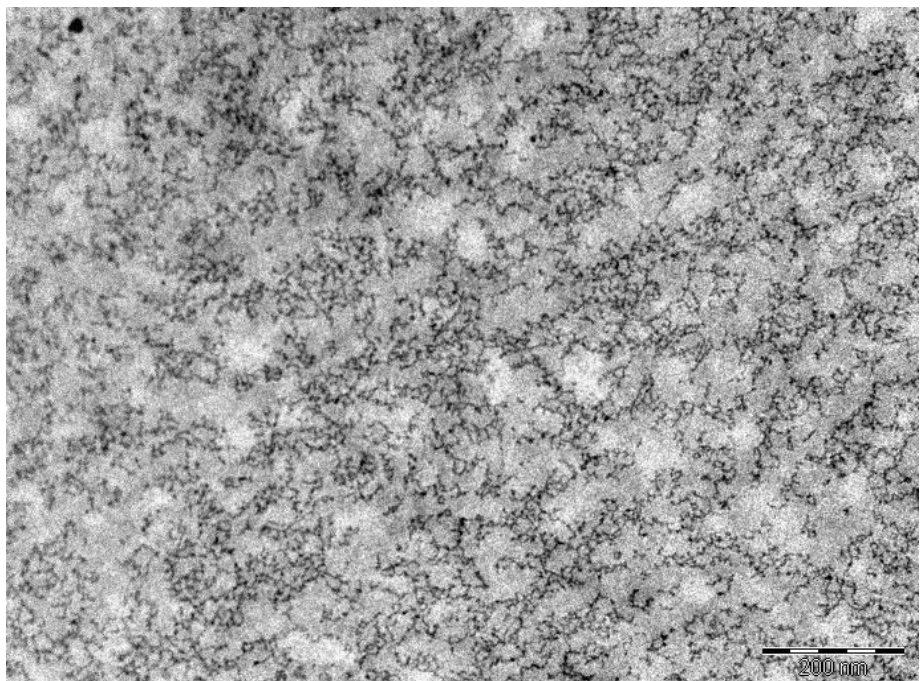

**Figure S3.** TEM image of the composite P3HT/PbS-Ol NCs 40:60 w/w.

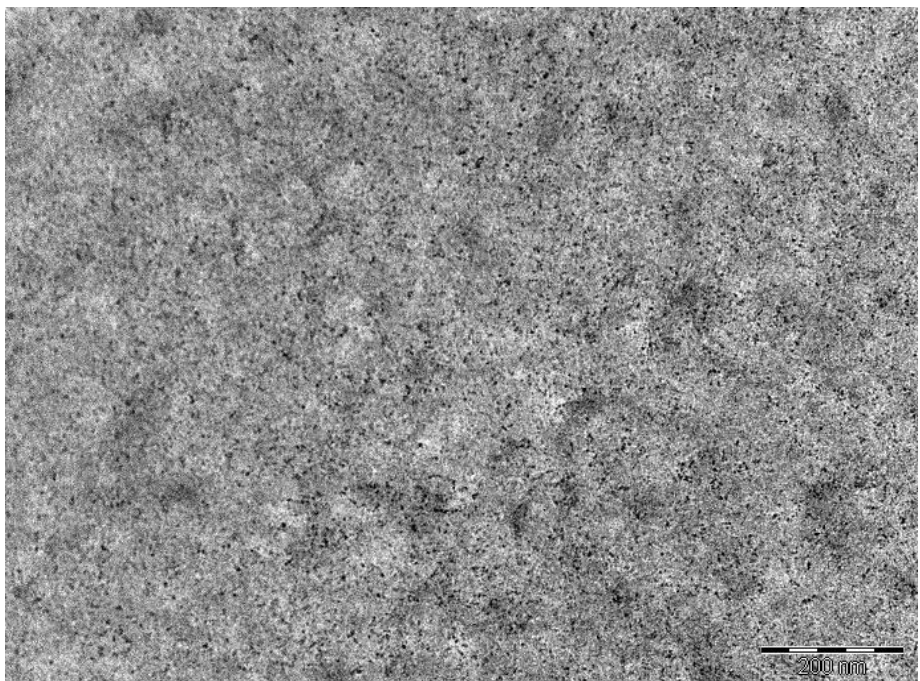

**Figure S4.** TEM image of the composite P3HT/PbS-Ol NCs 20:80 w/w.

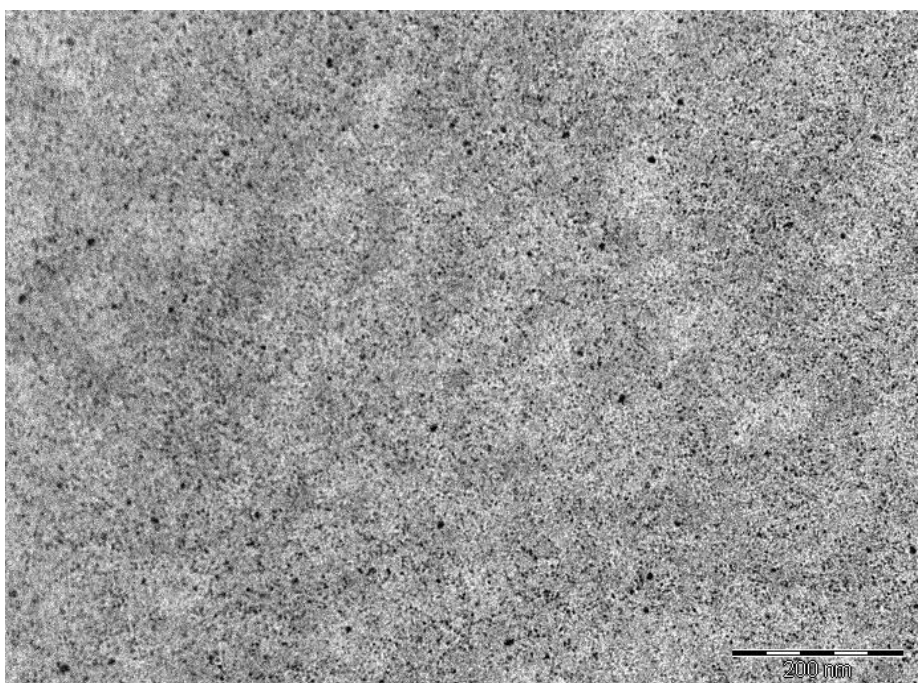

**Figure S5.** TEM image of the composite P3HT/PbS-Ol NCs 10:90 w/w.

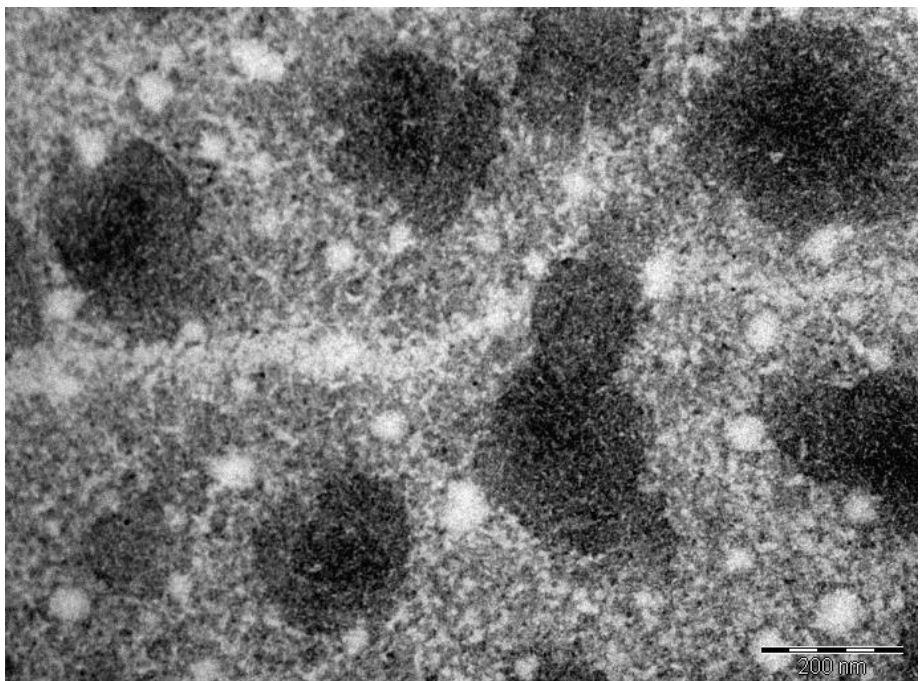

**Figure S6.** TEM image of the composite P3HT/PbS-Ol NCs 10:90 w/w, + MPA-treatment.

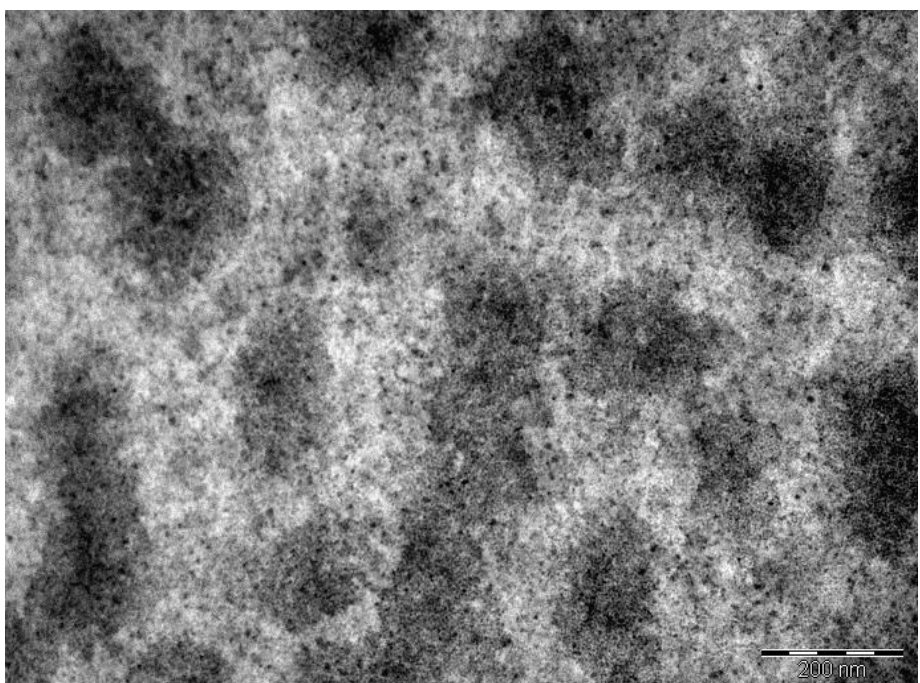

**Figure S7.** TEM image of the composite P3HT/PbS-Ol NCs 10:90 w/w, + annealing.

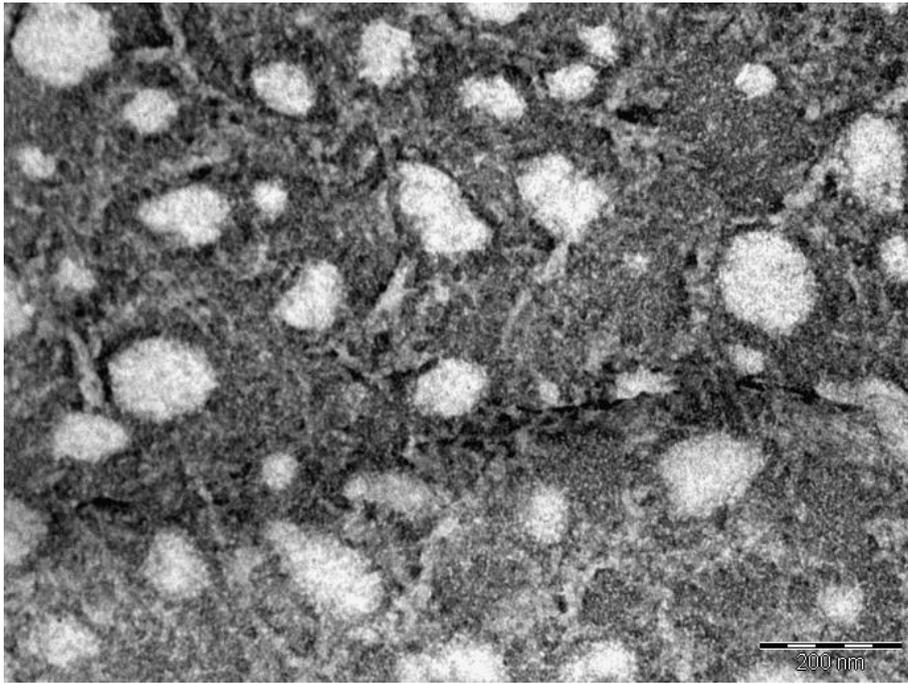

**Figure S8.** TEM image of the composite P3HT/PbS-Ol NCs 10:90 w/w, + MPA-treatment, + annealing.

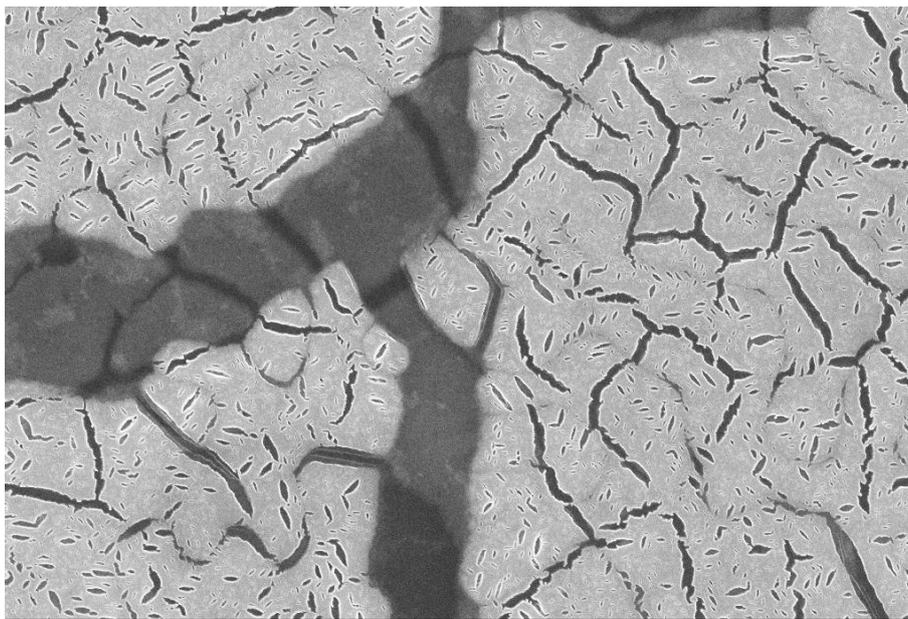

**Figure S9.** SEM image of the composite P3HT/PbS-Ol NCs 10:90 w/w, + MPA-treatment, + annealing. Scale bar is 2 μm.

*TEM images of P3HT/PbS-ArS NC composites*

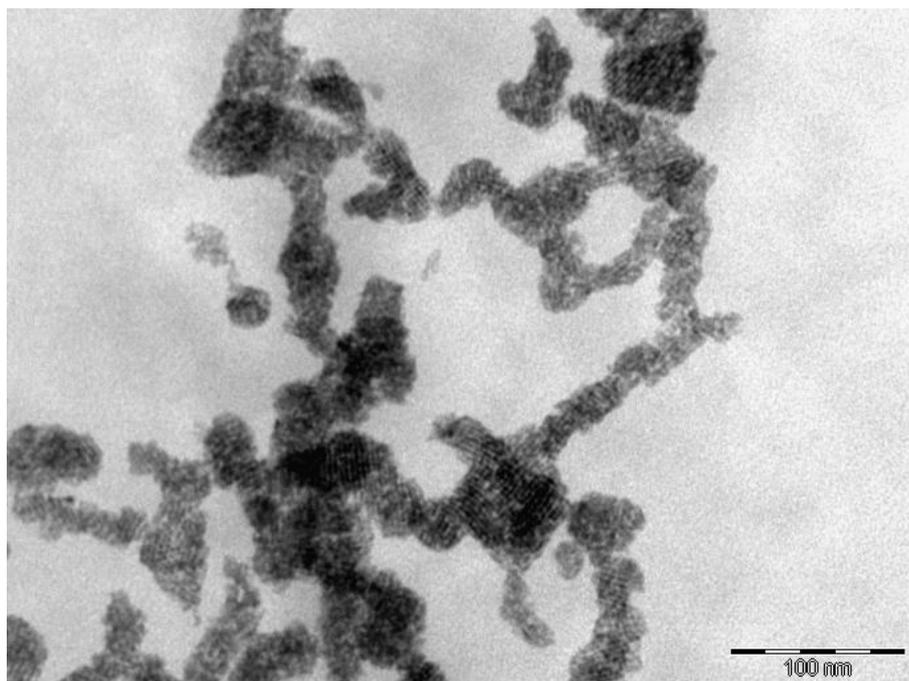

**Figure S10.** TEM image of the composite P3HT/PbS-ArS NCs 50:50 w/w.

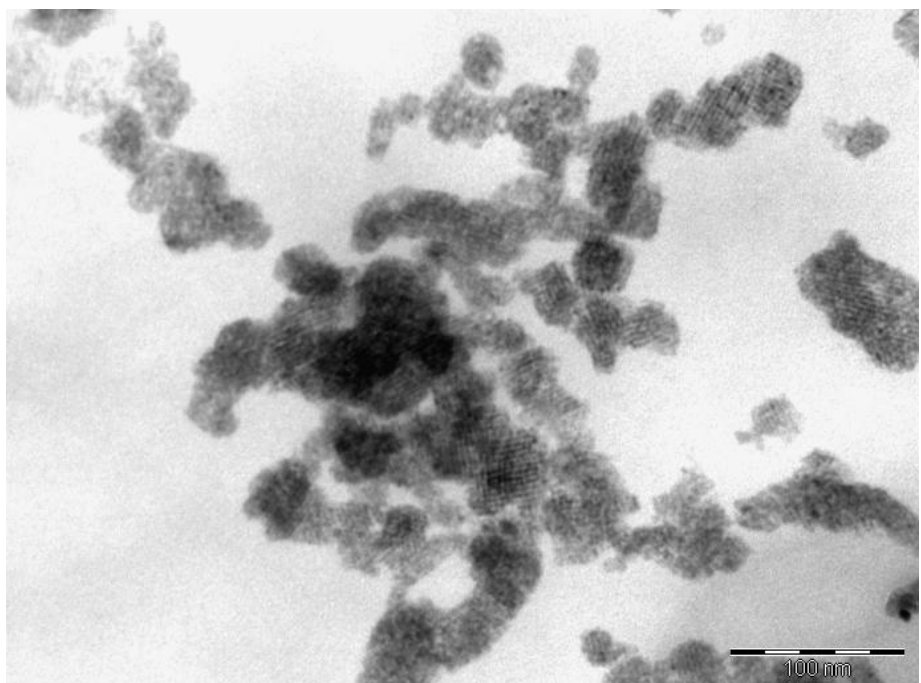

**Figure S11.** TEM image of the composite P3HT/PbS-ArS NCs 40:60 w/w.

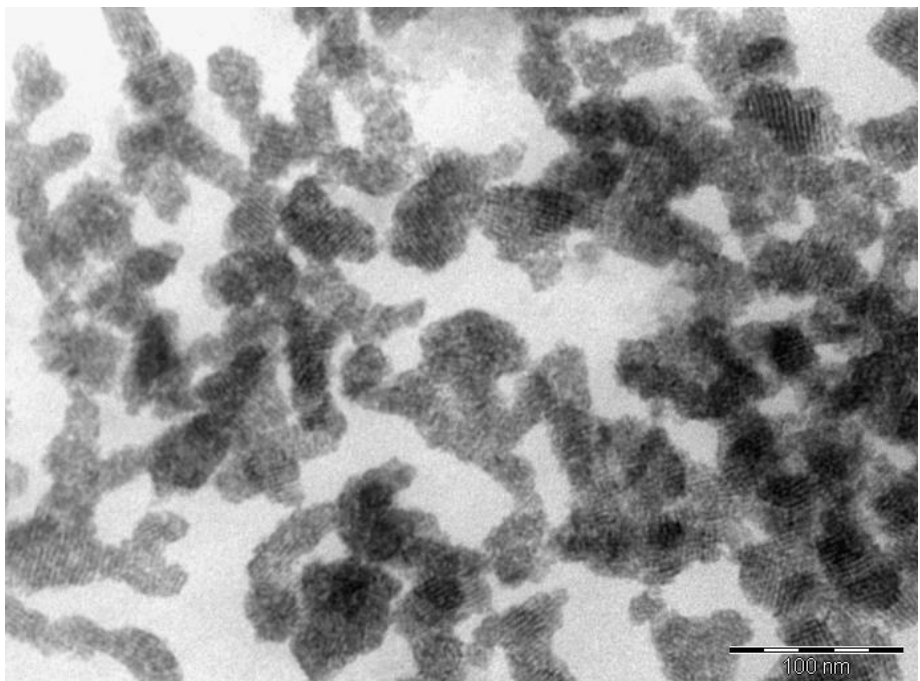

**Figure S12.** TEM image of the composite P3HT/PbS-ArS NCs 20:80 w/w.

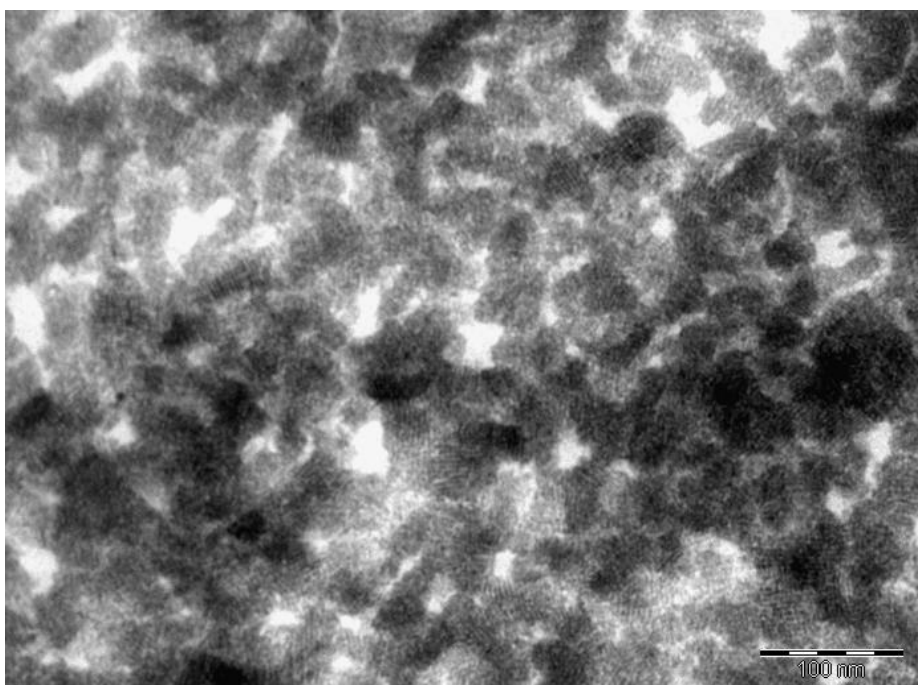

**Figure S13.** TEM image of the composite P3HT/PbS-ArS NCs 10:90 w/w.

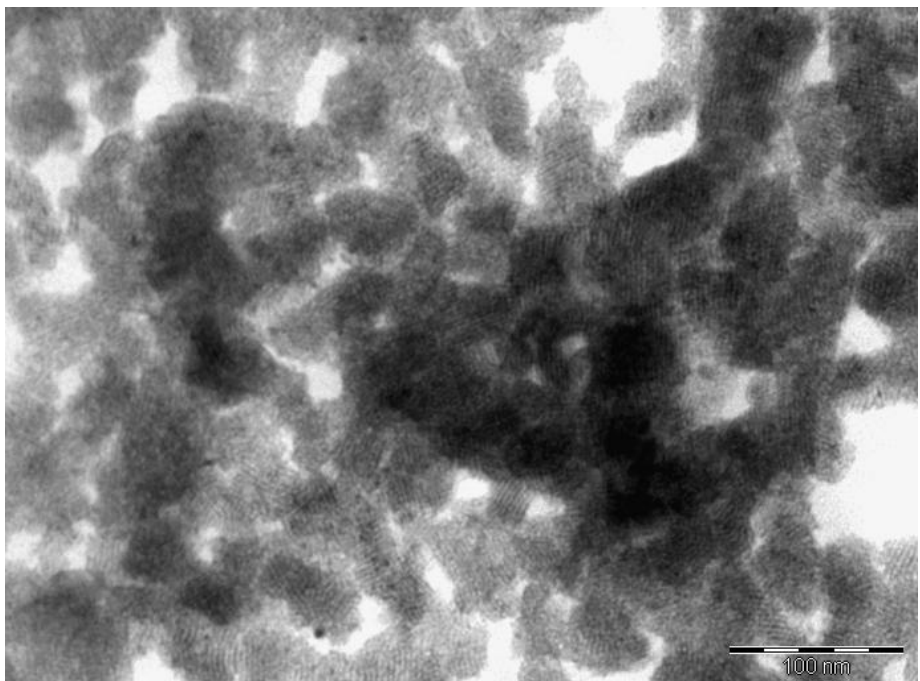

**Figure S14.** TEM image of the composite P3HT/PbS-ArS NCs 10:90 w/w, + MPA-treatment.

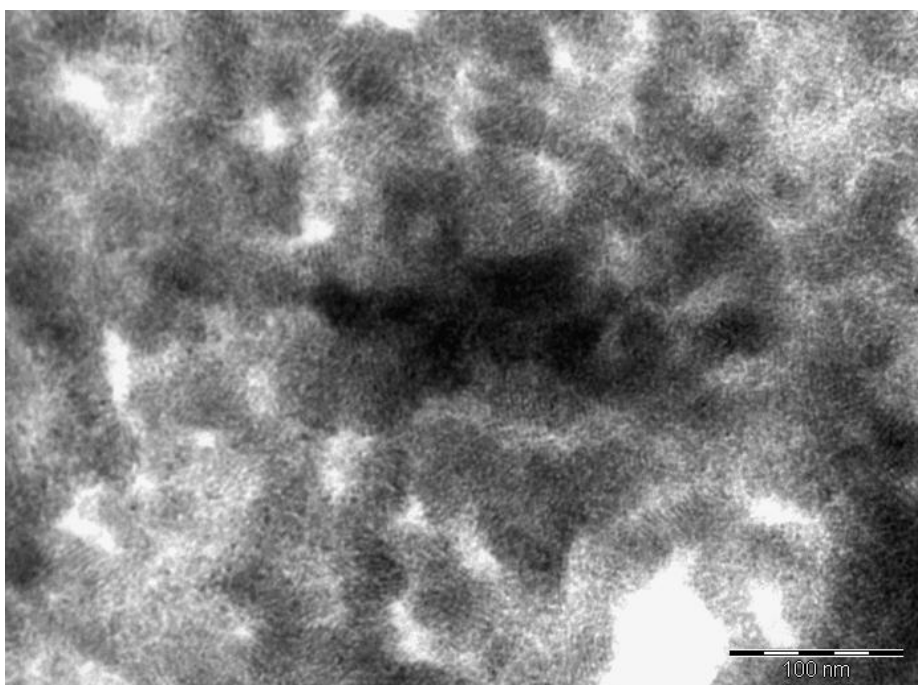

**Figure S15.** TEM image of the composite P3HT/PbS-ArS NCs 10:90 w/w, + annealing.

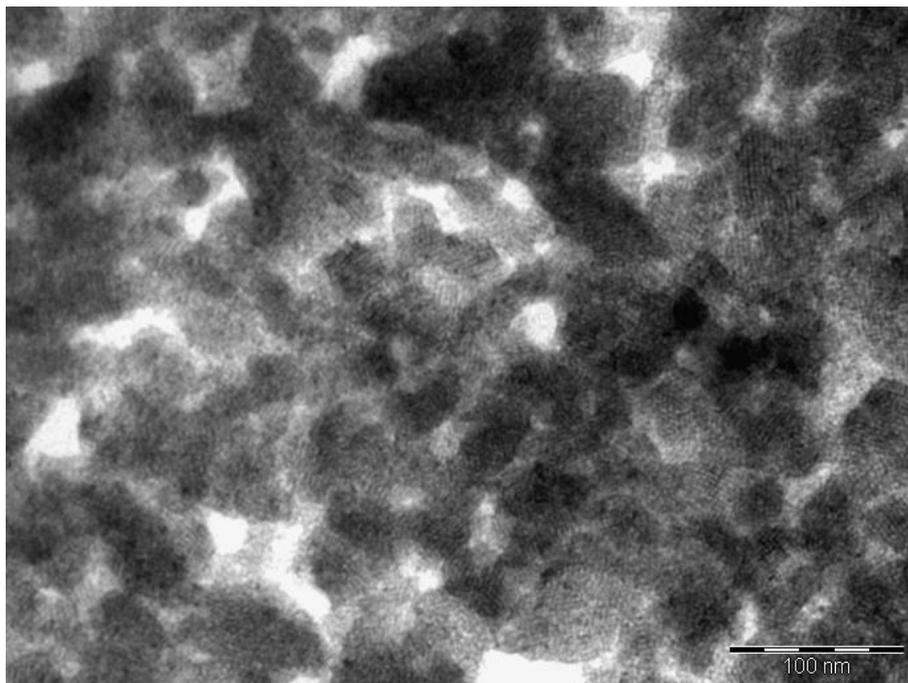

**Figure S16.** TEM image of the composite P3HT/PbS-ArS NCs 10:90 w/w, + MPA-treatment, + annealing.

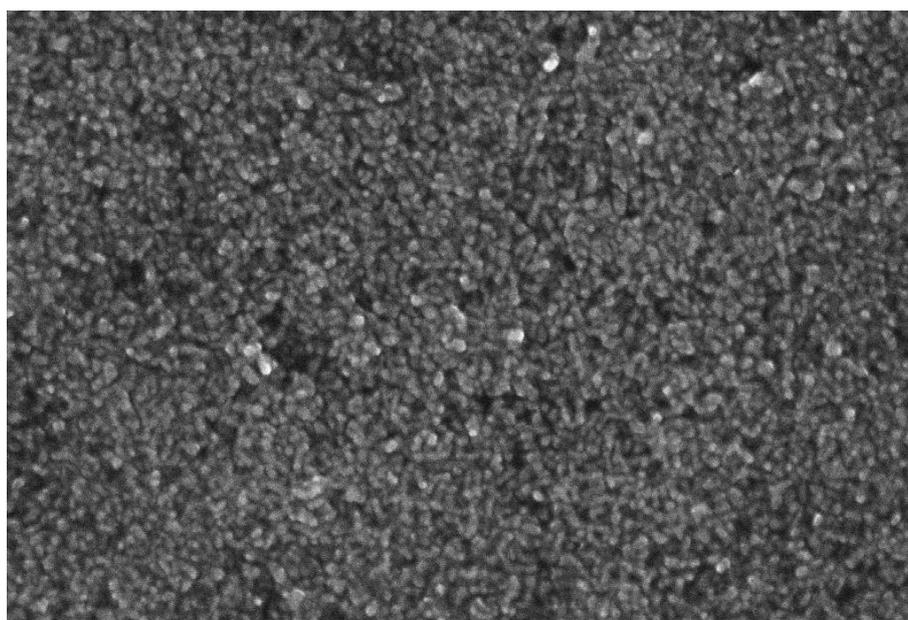

**Figure S17.** SEM image of the composite P3HT/PbS-ArS NCs 10:90 w/w, + MPA-treatment, + annealing. Scale bar is 200 nm.

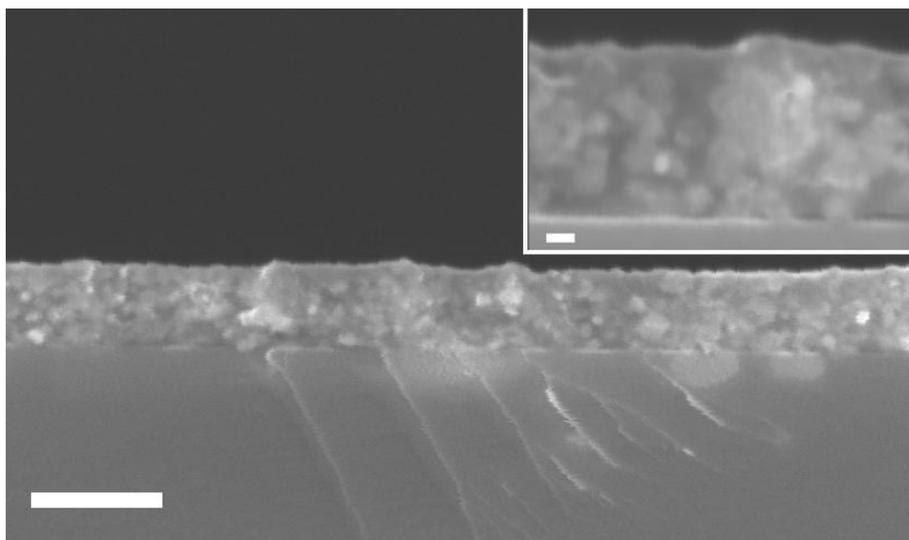

**Figure S18.** Cross section SEM image of the composite P3HT/PbS-ArS NCs 10:90 w/w, + MPA-treatment, + annealing; scale bar is 200 nm. Inset shows a magnified view of such composite (scale bar is 20 nm); darker and brighter zones can be distinguished corresponding to P3HT-rich and PbS NC-rich domains, respectively.

*AFM images of the nanocomposites*

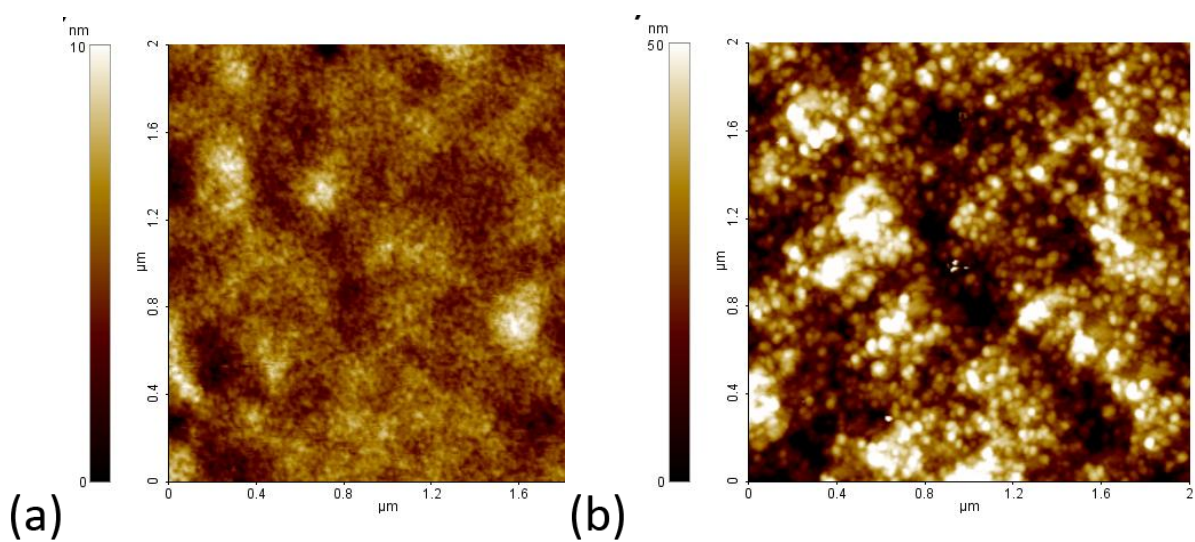

(a)          (b)

**Figure S19.** AFM images of the as-deposited hybrid composites: a) P3HT/PbS-Ol NCs 10:90 w/w, b) P3HT/PbS-ArS NCs 10:90 w/w.

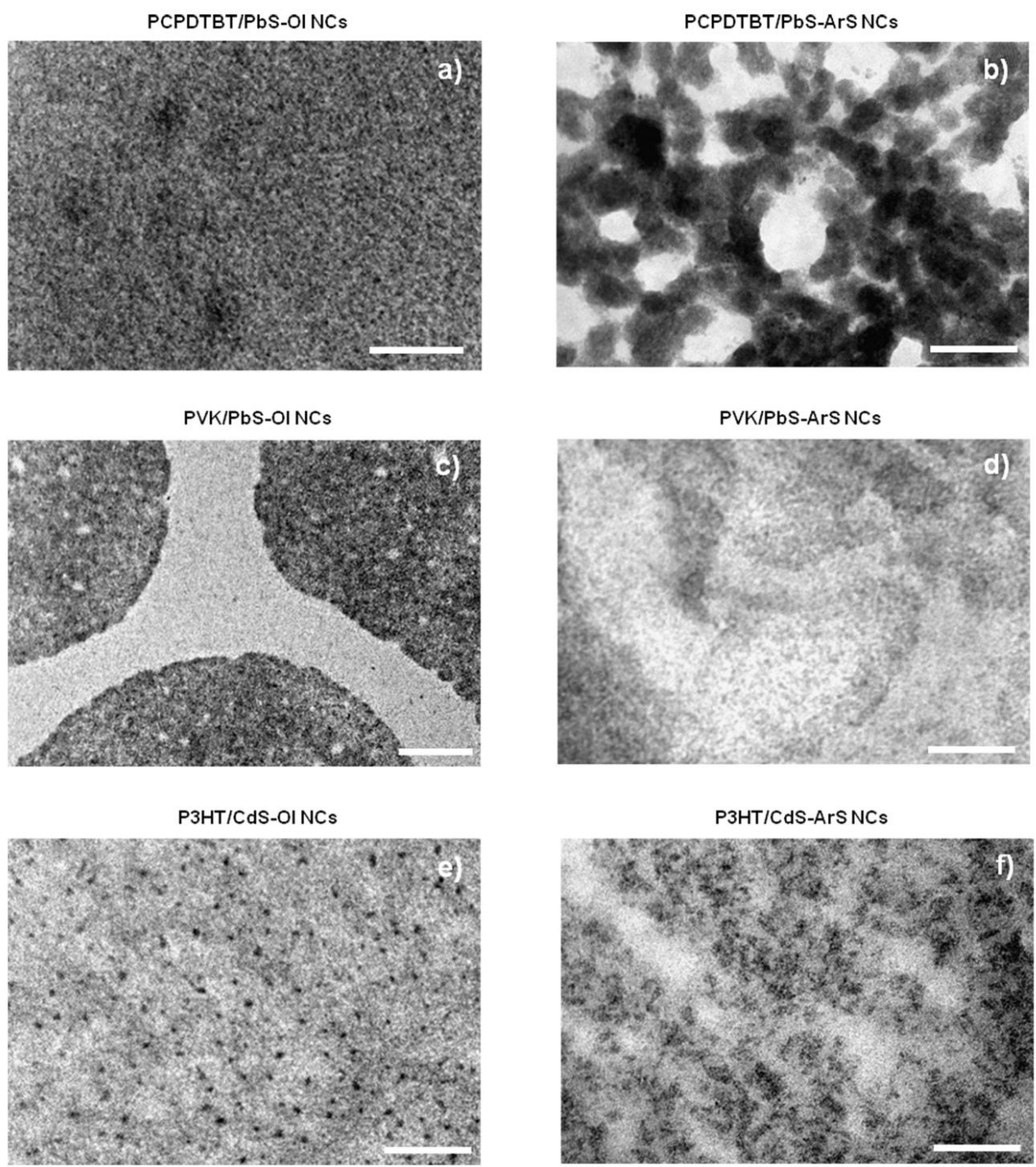

**Figure S20.** Transmission EM images of the as-deposited nanocomposites (the components constituting the composite represented are written on the top of the corresponding micrograph). Size scale bar appears in the bottom right corners and corresponds to 100 nm.

The general applicability of our approach has been demonstrated employing different CP and NC components. Indeed, PbS-Ol NCs are uniformly dispersed in a narrow band-gap polymer with alkyl side chains as Poly[2,6-

(4,4-bis-(2-ethylhexyl)-4H-cyclopenta[2,1-b;3,4-b′]dithiophene)-alt-4,7(2,1,3-benzothiadiazole)] (commonly known as PCPDTBT), while PbS-ArS NCs self-assemble in cubic close-packed nanometer size scale domains (see panels a and b of Figure S18, respectively). Further confirmation of the ligand-mediated control of CP/NC non-covalent interactions arises by blending PbS NCs with polystyrene poly(9-vinylcarbazole) (known as PVK) which is constituted by a saturated backbone with aromatic side groups: indeed, PbS-Ol NCs in PVK self-segregate in the PVK matrix, which is likely driven by aliphatic interactions between oleate ligands, whereas aromatic side chains of PVK may intercalate aryl pending moieties of PbS-ArS NCs guaranteeing their dispersion in the polymeric matrix (see panels c and d of Figure S18, respectively). Analogously, non-covalent interactions between P3HT and as-synthesized oleate-capped CdS NCs[3] (CdS-Ol NCs) or solution-phase ligand exchanged arenethiolate-capped CdS NCs (CdS-ArS NCs) lead to qualitatively similar morphologies to those observed for P3HT/PbS NC composites, in which CdS-Ol NCs are evenly dispersed, whereas CdS-ArS NCs show propensity to close-pack in nanoscale domains (see panels e and f of Figure S18, respectively).

*Device fabrication and characterization.*

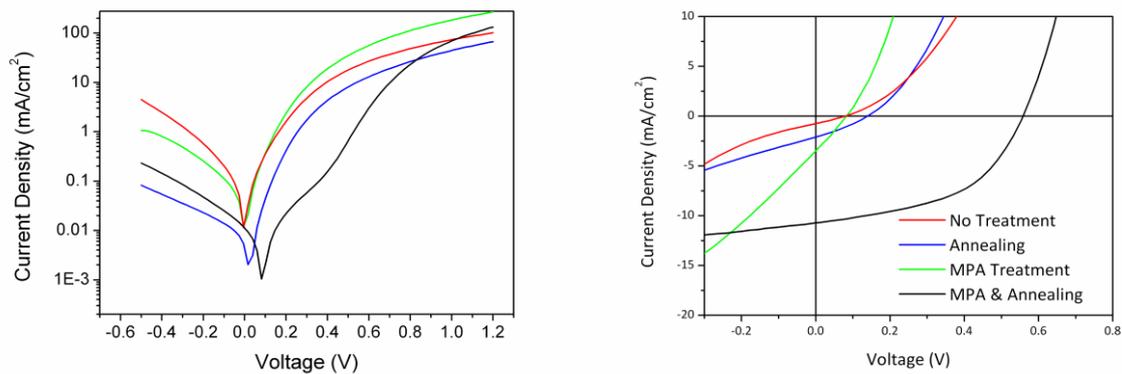

**Figure S21.** *i*-V characteristics showing the effect of device post-deposition treatments.

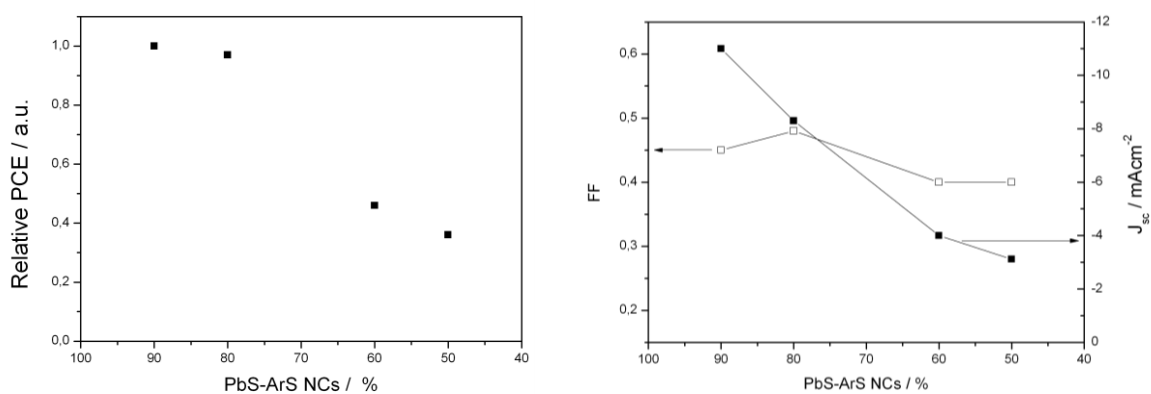

**Figure S22.** Relative PCE, FF and J$_{sc}$ values for devices based P3HT/PbS-ArS NCs at different weight ratio.

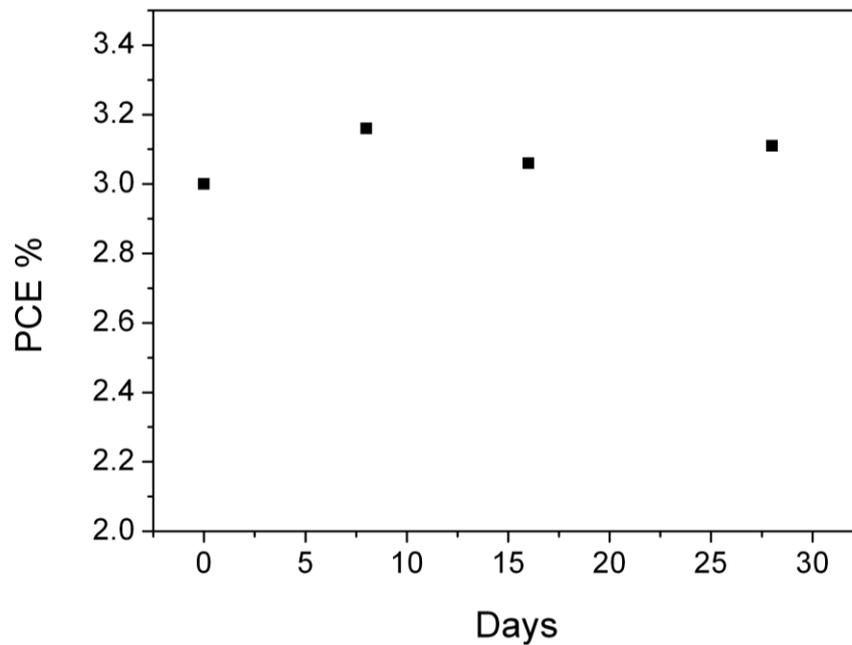

**Figure S23.** Temporal evolution of the PCE of device comprising P3HT/PbS-ArS NC composite stored in inert atmosphere.

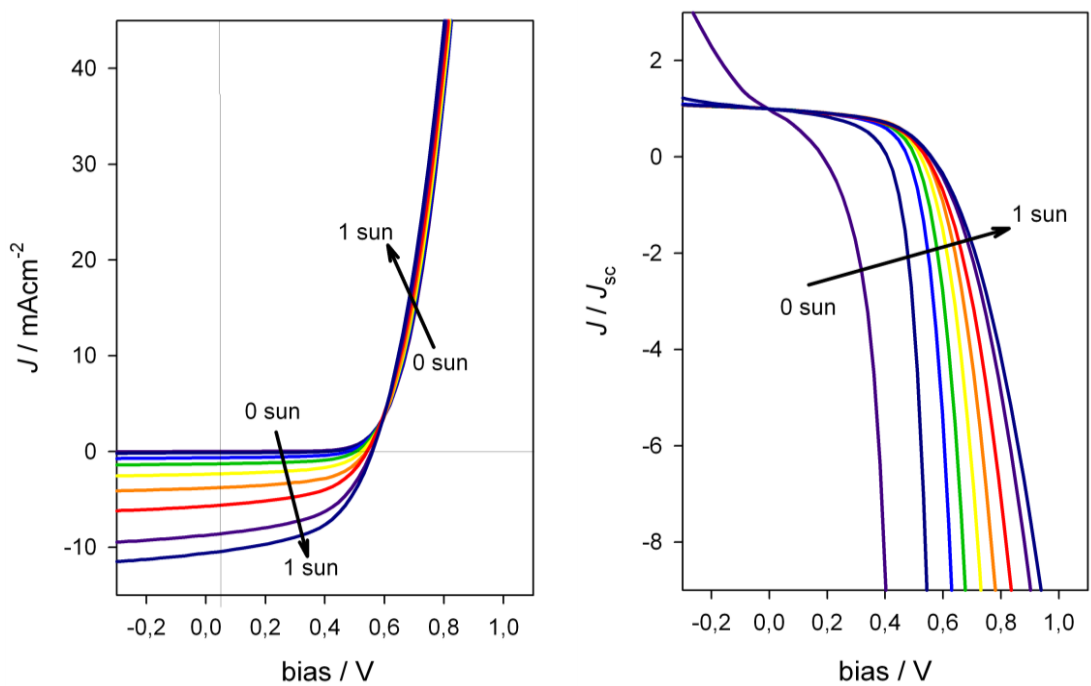

**Figure S24.** *i*-V characteristics showing the effect of incident light power density on device comprising P3HT/PbS-ArS NC composite.

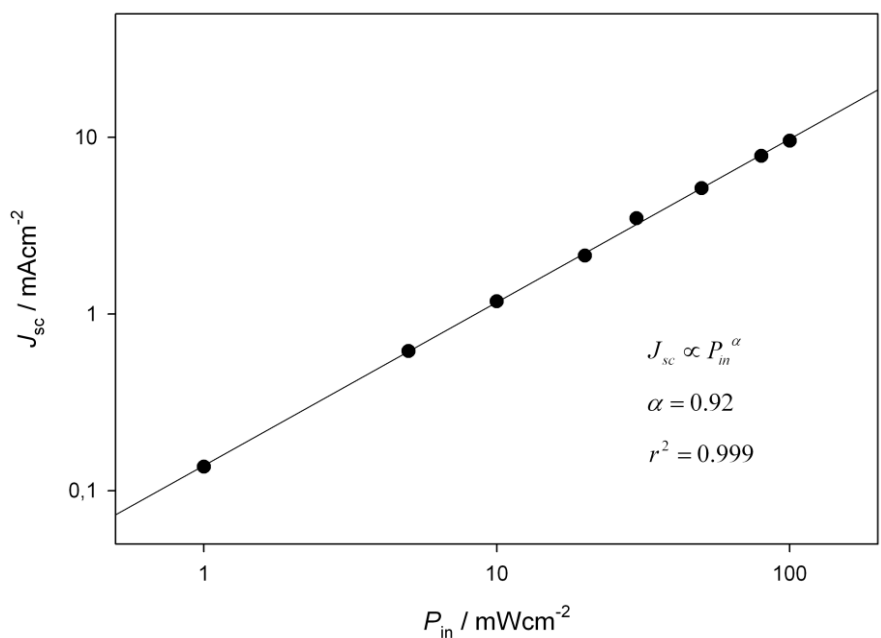

**Figure S25.** Log-log plot of $J_{sc}$ vs. $P_{in}$ for device comprising P3HT/PbS-ArS NC composite.

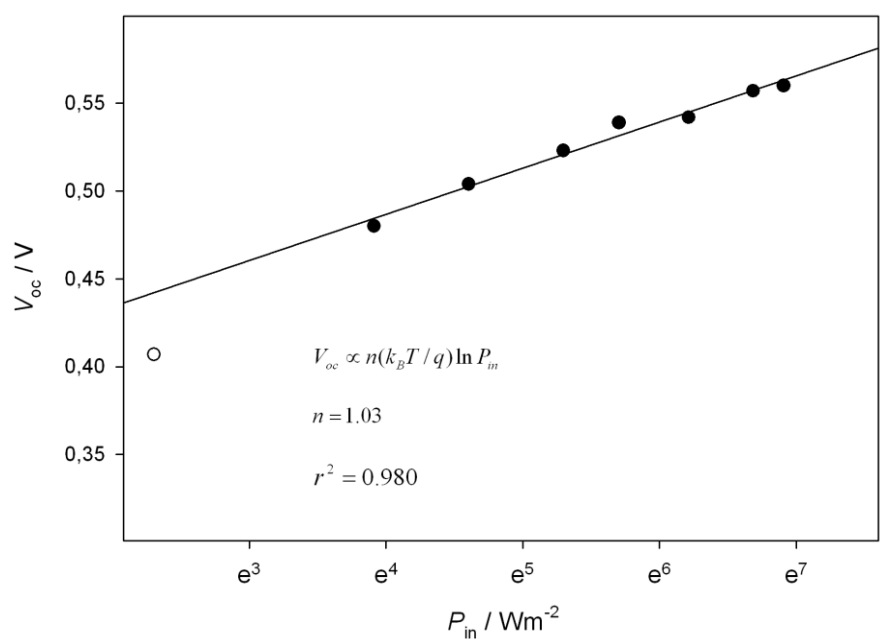

**Figure S26.** Semi-ln plot of $V_{oc}$ vs. $P_{in}$ for device comprising P3HT/PbS-ArS NC composite.

*Scanning Kelvin Probe Microscopy measurements of P3HT/PbS NC composites and their components*

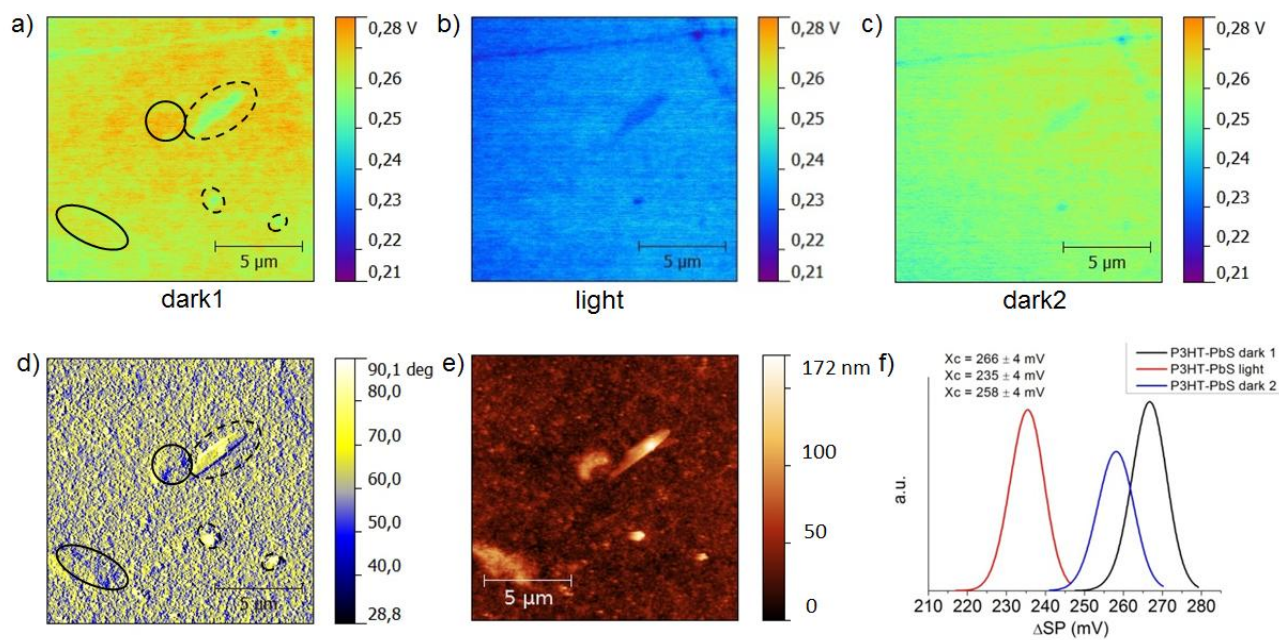

**Figure S27.** Scanning Kelvin Probe Force Microscopy Characterization. Surface Potential maps of the P3HT/PbS-ArS NC composite (a) before illumination (dark1), (b) under illumination (light) with a white light source and (c) after illumination (dark2). (d) Phase and (e) topography AFM images of the P3HT/PbS-ArS composite. (f) Histogram representation of the surface potential images, highlighting the potential shift under illumination.

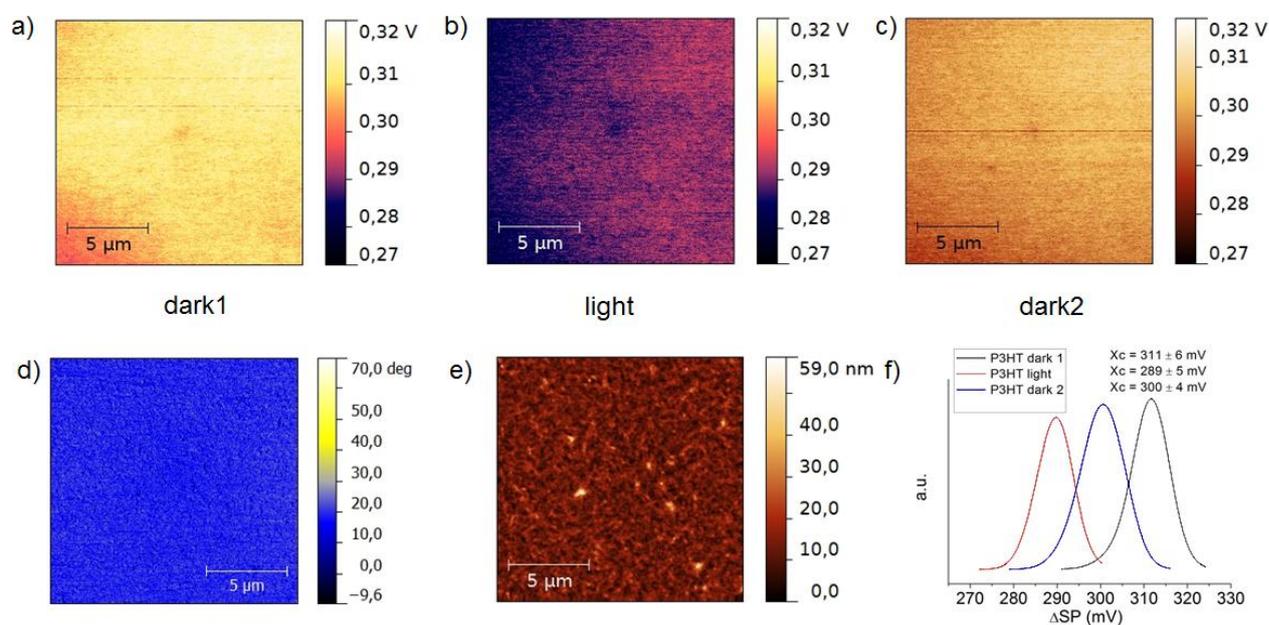

**Figure S28.** Surface Potential mappings of the P3HT pristine film (a) before illumination (dark1), (b) under illumination (light) with a white light source and (c) after illumination (dark2). (d) Phase and (e) topography AFM images of the P3HT/PbS-ArS composite. (f) Histogram representation of the surface potential images, highlighting the potential shift under illumination.

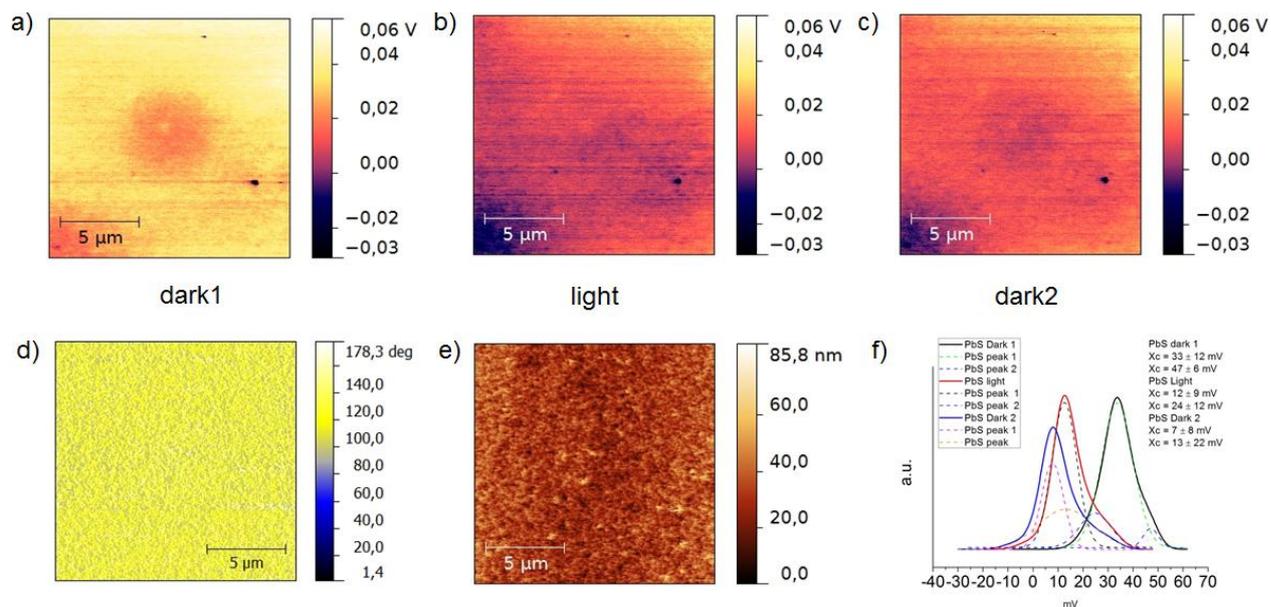

**Figure S29.** Surface Potential mappings of the PbS-ArS pristine film (a) before illumination (dark1), (b) under illumination (light) with a white light source and (c) after illumination (dark2). (d) Phase and (e) topography AFM images of the P3HT/PbS-ArS composite. (f) Histogram representation of the surface potential images, highlighting the potential shift under illumination.

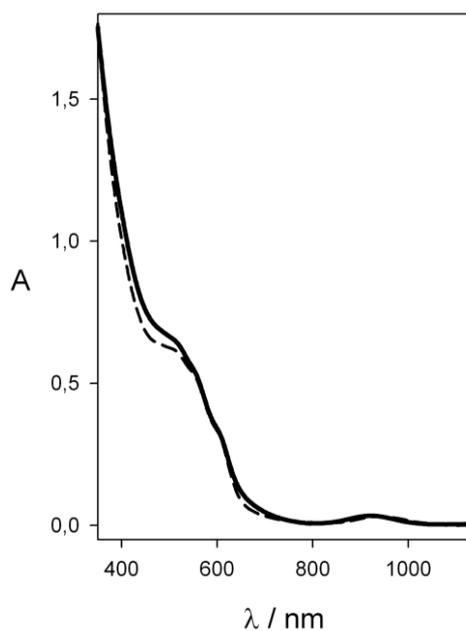

**Figure S30.** Absorption spectrum of the P3HT/PbS-ArS NC composite (solid line) and sum spectrum of pure P3HT and PbS-ArS NC solid (dashed line).

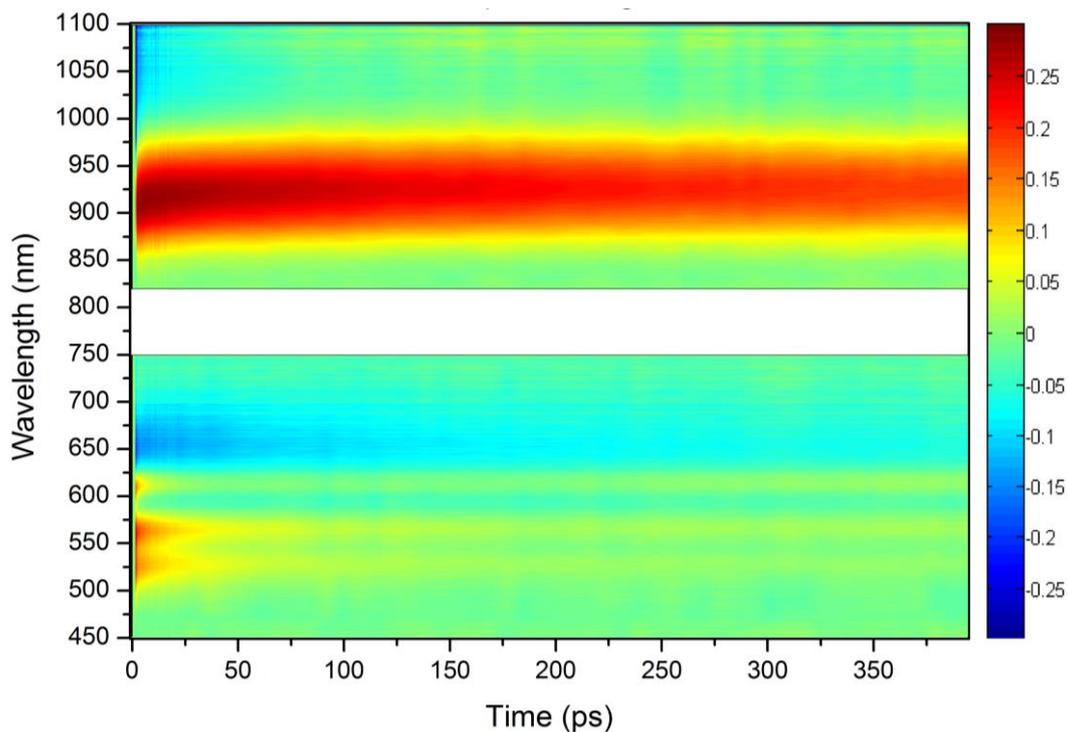

**Figure S31.** TA spectrum of P3HT/PbS-ArS NC composite ($\lambda_{pump}$ = 400 nm at a pulse irradiance of 20 µJ/cm$^2$).

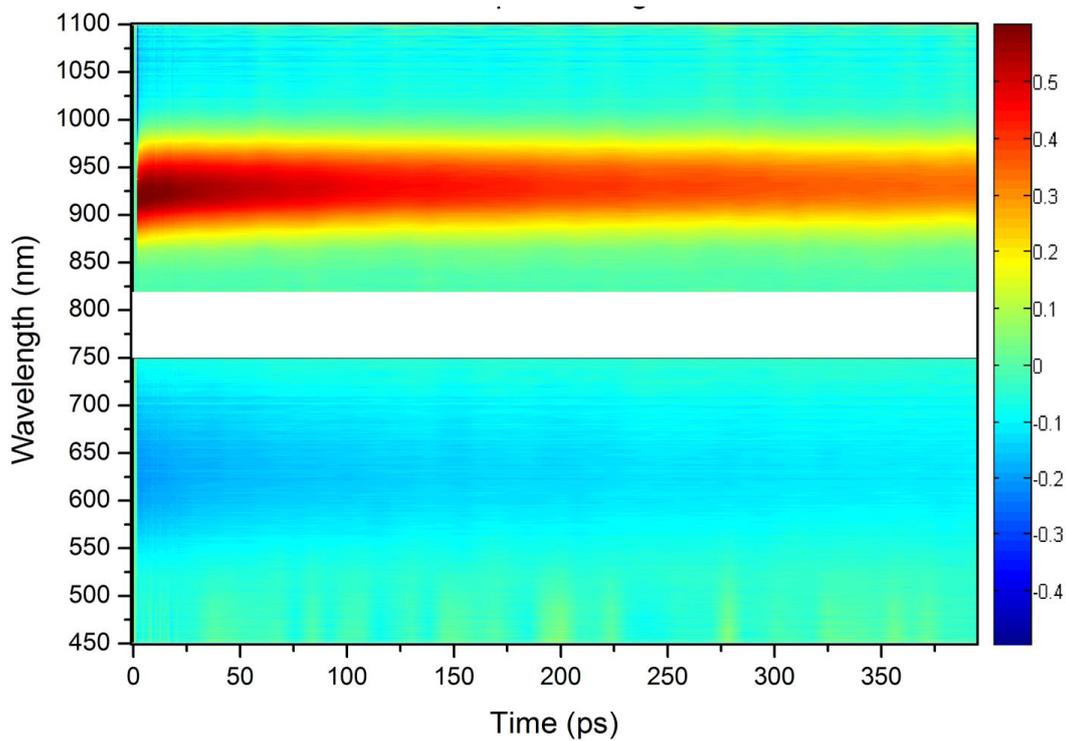

**Figure S32.** TA spectrum of PbS-ArS NC solid ($\lambda_{pump}$ = 400 nm at a pulse irradiance of 20 μJ/cm$^2$).

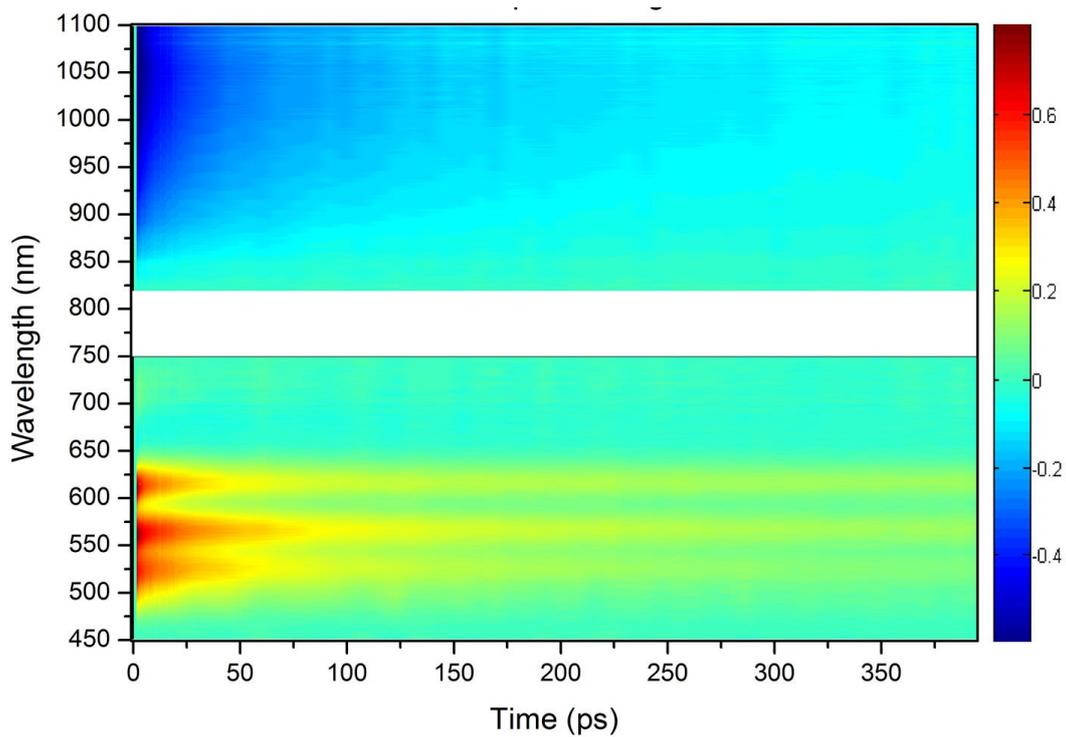

**Figure S33.** TA spectrum of pure P3HT ($\lambda_{pump}$ = 400 nm at a pulse irradiance of 20 μJ/cm$^2$).

**Supporting References.**

1.  Hines, M. A.; Scholes, G. D., Colloidal PbS Nanocrystals with Size-Tunable Near-Infrared Emission: Observation of Post-Synthesis Self-Narrowing of the Particle Size Distribution. *Adv. Mater.* **2003,** *15* (21), 1844-1849.

2.  Giansante, C.; Carbone, L.; Giannini, C.; Altamura, D.; Ameer, Z.; Maruccio, G.; Loiudice, A.; Belviso, M. R.; Cozzoli, P. D.; Rizzo, A.; Gigli, G., Colloidal Arenethiolate-Capped PbS Quantum Dots: Optoelectronic Properties, Self-Assembly, and Application in Solution-Cast Photovoltaics. *J. Phys. Chem. C* **2013,** *117* (25), 13305-13317.

3.  Yu, W. W.; Peng, X., Formation of High-Quality CdS and Other II–VI Semiconductor Nanocrystals in Noncoordinating Solvents: Tunable Reactivity of Monomers, *Angew. Chem. Int. Ed.* **2002,** *41* (13), 2368-2371.